# STRUCTURAL CHARACTERISTICS AND STELLAR COMPOSITION OF LOW SURFACE BRIGHTNESS DISK GALAXIES

STACY S. MCGAUGH

Department of Astronomy, University of Michigan, Ann Arbor, MI 48109

and

Institute of Astronomy, University of Cambridge,

Madingley Road, Cambridge CB3 0HA, England[1]

I: ssm@mail.ast.cam.ac.uk

AND

GREGORY D. BOTHUN

Physics Department, University of Oregon, Eugene, OR 97403

I: nuts@moo.uoregon.edu

[1]Current address

# ABSTRACT

We present $UBVI$ surface photometry of a sample of low surface brightness (LSB) disk galaxies. LSB disk galaxies are fairly well described as exponential disks with no preferred value for either scale length, central surface brightness, or rotational velocity. Indeed, the distribution of scale lengths is indistinguishable from that of high surface brightness spirals, indicating that dynamically similar galaxies (e.g., those with comparable $v^2 R$) exist over a large range in surface density.

These LSB galaxies are strikingly blue. The complete lack of correlation between central surface brightness and color rules out any fading scenario. Similarly, the oxygen abundances inferred from H II region spectra are uncorrelated with color so the low metallicities are not the primary cause of the blue colors. While these are difficult to interpret in the absence of significant star formation, the most plausible scenario is a stellar population with a young mean age stemming from late formation and subsequent slow evolution.

These properties suggest that LSB disks formed from low initial overdensities with correspondingly late collapse times.

*Subject headings:* galaxies: evolution — galaxies: formation — galaxies: stellar content — galaxies: structure

# 1. INTRODUCTION

Our knowledge of the physical properties of spiral galaxies is based on observations of high surface brightness (HSB) disks which are selected primarily for their striking appearance. In contrast, low surface brightness (LSB) galaxies are notoriously difficult to identify and study in detail. With peak surface brightnesses at or below the level of the sky background, it is not surprising that little data of this sort exists (e.g., Romanishin, Strom, & Strom 1983). There is no guarantee that the HSB spirals which define the Hubble Sequence adequately represent all types of disk galaxies, and there is good reason to suspect that they represent only a narrow range of the physical parameter space occupied by galaxies (e.g., Disney 1976). Here, we investigate the properties of a collection of LSB galaxies in order to extend our physical knowledge of disks to lower surface brightnesses.

The galaxies discussed here are drawn from either the UGC (Nilson 1973), or the lists of LSB galaxies compiled by Schombert & Bothun (1988) and Schombert et al. (1992). Galaxies are selected to have central surface brightnesses $\mu_0 \geq 23$ mag arcsec$^{-2}$, which we take as an operative definition of low surface brightness. Of the many candidates which meet this qualification, the sample selected for detailed photometric study was chosen to match that for which H II region spectroscopy was being obtained (McGaugh 1993) so that metallicity information is available to supplement the interpretation of the photometry. Though less numerous than in HSB spirals, H II regions are not uncommon in LSB disks (McGaugh 1992). LSB galaxies generally contain substantial amounts of atomic hydrogen (Schombert et al. 1992), but exhibit a deficit of molecular gas (Schombert et al. 1990). It is unclear whether or not the H II regions are associated with molecular cloud complexes at all, and we are not witnessing the kind of spiral density wave induced formation of OB associations in giant molecular clouds that characterizes star formation in Hubble Sequence spirals.

Despite the paucity of molecular gas, the large amounts of neutral hydrogen indicate substantial gas mass fractions (McGaugh et al. 1993). Our sample thus contains gas rich field disks with some ongoing star formation. The sample spans a wide range in redshift, from $cz = 1000$ km s$^{-1}$ to 12000 km s$^{-1}$. The median is $\sim 4000$ km s$^{-1}$, comparable to that of an HSB sample selected with the same angular diameter criteria (Schombert et al. 1992). While our sample is not complete in any sense nor exhaustive of all possible types of LSB galaxies, it fairly represents the types of objects found in current surveys for LSB galaxies, and provides the opportunity to examine the physical properties of some examples of a neglected class of galaxies.

Distances are derived from the recession velocity assuming $H_0 = 100\,h^{-1}\mathrm{km\,s^{-1}Mpc^{-1}}$ with $h = 1$ and a Virgocentric infall velocity of 300 km s$^{-1}$. For those objects with $cz < 3000$ km s$^{-1}$, the distance was found by interpolating through Table 3 of Aaronson et al. (1982). These indicative distances provide a basis for the comparison of absolute quantities. The adopted short distance scale provides conservative lower limits on the sizes and luminosities of these LSB systems.

## 2. OBSERVATIONS AND REDUCTIONS

Images of 20 LSB galaxies were obtained through $UBVI$ filters with the McGraw Hill 1.3 m telescope of the MDM[1] Observatory in February 1991, November 1991, and January 1992. During the February 1991 observing session, a TI-4849 CCD (Luppino 1989) was employed as the camera. For the other runs, a Thomson CCD was used. Both of these devices gave essentially the same scale, 0.48 arcsec pixel$^{-1}$ (Aldering 1990).

Initial reductions followed standard procedures, and were performed with IRAF[2]. This involved bias and dark current subtraction, and flat fielding. For the Thomson CCD, the bias level for each frame was taken from the overscan region. The dark current of the Thomson CCD was fairly steady at $\sim 18\,e^-$pixel$^{-1}$hr$^{-1}$. Because of the many low level charge transfer traps apparent in long dark current measurements, this number was subtracted as a constant from each frame (rather than subtracting the cosmetically poor dark frame from which this number was derived). The traps were completely filled by counts from the sky in the broad band exposures. For the TI-4849, no record is kept of the overscan level, so a large number ($> 20$) of bias frames were used to determine the zero level. Unlike the Thomson, this CCD exhibited structure in its dark current, including a large number of hot pixels and a low frequency wave pattern. These were removed by the subtraction of a dark frame which was the median of many, effectively removing cosmic rays.

Flat fields were determined from exposures of the twilight sky, or of mostly blank fields of the dark sky if time allowed. In general, the flat fields are determined to 0.1% in $B$, 0.2% in $U$ and $V$, and 0.5% in $I$. A small residual gradient is present in the I, but neither CCD suffers from fringing. Typical exposure times were 3600 seconds in $U$, 1800 in $B$, and 900 in $V$ and $I$.

---

[1] MDM Observatory is operated by the University of Michigan, Dartmouth College, and the Massachusetts Institute of Technology.

[2] The Image Reduction and Analysis Facility (IRAF) is distributed by the Association of Universities for Research in Astronomy, Inc., under contract to the National Science Foundation.

The data were calibrated with observations of standard stars from the lists of Landolt (1973, 1983). As a result of the eruption of Mt. Pinatubo, the extinction coefficients for the November 1991 observing run were found to depart noticeably from the average values for Kitt Peak, consistent with the values reported by Pilachowski et al. (1991). These were assumed to apply to the January 1992 data as well, as bad weather prevented acquisition of enough calibration data to accurately measure the extinction coefficients during that run. Sufficient data were obtained to determine the zero point during photometric periods; these are accurate to $\lesssim 0.1$ mag. The color solutions from the two runs using the same filters and detector are identical within the errors. Colors are accurate to $\sim 0.05$ mag.

### 3. ANALYSIS AND RESULTS

#### 3.1 *Structure*

Surface photometry was performed on the images following the methods described by Bothun et al. (1986). The best surface photometry comes form the $B$-band frames as these are the deepest and flattest. Typically, over 3 scale lengths of disk light are contained in the $B$ images. Ellipses were fit allowing the position angle, ellipticity and center to vary. If necessary, bulge and disk components were deconvolved from the surface brightness profile using the procedure outlined by Schombert & Bothun (1987). Most, though certainly not all, of these galaxies lack significant bulge components. If possible, an exponential profile of the form

$$\Sigma(r) = \Sigma_0 \, e^{-r/\alpha} \qquad (1)$$

was fit to the disk component of each galaxy. Here, $\Sigma$ is the surface luminosity density in linear units (e.g., $L_\odot \, \text{pc}^{-2}$), corresponding to the surface brightness $\mu$ (in mag arcsec$^{-2}$), and $r$ is the major axis radius. The fit parameters are then the central surface brightness $\Sigma_0$ ($\mu_0$) and exponential scale length $\alpha$. As is the case for HSB spirals, exponential profiles generally provide an adequate description of the radial distribution of the light in LSB galaxies, and are the best objective descriptor of their structure.

The radial surface brightness distributions and exponential fits are presented in Figure 1. The structural parameters ($\mu_0, \alpha$) derived from the fits are listed in Table 1, together with other relevant information. Column 1 contains the name of the galaxy. Those objects starting with 'F' are from the lists of Schombert & Bothun (1988) or Schombert et al. (1992), while those designated by 'U' are from the UGC. UGC 12695 has two entries, corresponding to the two acceptable fits (see Figure 1). Column 2 gives the major axis diameter at the 25 mag arcsec$^{-2}$ level (in arcseconds). Columns 3 and 4 give the fit parameters $\alpha$ and $\mu_0$. The former is in arcseconds and the latter in $B$ mag arcsec$^{-2}$. The central surface brightness has been corrected for galactic extinction using the $A_B$ values of Burstein & Heiles (1984). It has not been corrected for cosmological $(1+z)^4$ dimming, which is negligible in most cases. Formally, the errors on these quantities are small, $\mu_0$ in principle being limited only by the uncertainty in the zero point. However, the fits are subject to systematic errors in the placement of the sky level, and the scale length will depend on the filter if color gradients are present. Also, some of the profiles in Figure 1 show an indication of a central concentration which is not well fit as a bulge, but might be described as a short scale length exponential star pile (Kormendy 1993). Such difficulties and the basic limitation on the degree to which the exponential model is appropriate limits the accuracy of the fit parameters to $\sim 20\%$. For a few objects in common, van der Hulst & de Blok (1993) obtain similar results from somewhat deeper images.

The exponential scale length (in kpc) is given in column 5 for the (short) adopted distance (column 10). This is the best indicator of the size of LSB disks, as isophotal diameters will always provide a systematic underestimate. That is, for two disks with the same scale length but different central surface brightnesses, the measured isophotal diameter will obviously be less for the lower surface brightness disk simply because any given isophote is reached at a smaller radius owing to the fainter start.

This is illustrated by the difference between the isophotal and total magnitudes for LSB disks. The sixth and seventh columns list the apparent $B$ magnitudes as measured within the 25 mag arcsec$^{-2}$ isophote (column 6) and by the integration of the exponential profile (column 7). The latter comes from the integration of equation (1) over the area of the disk, which gives

$$L = 2\pi \alpha^2 \Sigma_0. \qquad (2)$$

Figure 2 demonstrates the severity of the underestimate of the amount of light produced by LSB galaxies when an isophotal magnitude is used. Hence, we will use equation (2) to compute and compare disk luminosities. The "total" quantities in Table 1 refer to the disk component *only*. The total disk absolute magnitude is listed in column 8. All magnitudes have been corrected for extinction within our own galaxy, as with $\mu_0$.

The inclination is derived from the axial ratios of the ellipses fit at large radii, and is listed in column 9. These are computed from the Holmberg formula assuming that the disks are truly circular:

$$cos^2(i) = \frac{(b/a)^2 - q_0^2}{1 - q_0^2}, \qquad (3)$$

where $a$ and $b$ are the major and minor axes and $q_0$ is the intrinsic axial ratio of an edge-on system. This latter is unknown but probably small in the case of LSB galaxies; we adopt $q_0 = 0.1$. The assumed value of $q_0$ is unimportant, as its effect is small compared to the uncertainty in $i$. The LSB nature of these galaxies means that there is a considerable amount of uncertainty in the parameters of the ellipses fit at large radii. While the fit to the surface brightness profile is not very sensitive to this, the derived inclination is rather more susceptible. In most cases, the scatter in the ellipticities indicates an uncertainty in the inclination of typically $\sim 10°$, though it is considerably better in a few cases and worse in others.

### 3.2 Colors

Table 2 contains the colors of those galaxies observed through all four $UBVI$ filters. Each color has three entries. The first is the color of the inner regions as measured through a small ($10''$) aperture. In some cases, this color represents a true color of the bulge component while in others it reflects the presence of a somewhat HSB nucleus. The second is the luminosity weighted color, i.e., the color measured through a large isophotal aperture. This is most appropriate for comparison to data in the literature for which no decomposition of the bulge and disk components has been made. The third entry is the area weighted color. This is determined from the average of the colors of many smaller apertures over the entirety of the disk. These are either the rings in the differential color profile, or a grid of squares covering the galaxy ("two color mapping," Bothun 1986). Sometimes the signal in the $I$ band was too weak for this subdivision to be successful, so only the large aperture color is given for $V - I$. In principle, we feel that the area weighted colors should be more representative of the disk population. However, the luminosity and area weighting procedures give generally similar results as the bulge is usually very weak (though sometimes entirely dominant) over the observed area.

A few galaxies such as UGC 12695 have very bright H II regions. These may skew the luminosity weighted colors towards those of the H II regions. These can have peculiar colors because of the complex, rapidly evolving population of massive stars (Campbell & Terlevich 1984, McGaugh 1991), and because of emission line contamination (Salzer et al. 1991). The area weighted colors are less affected and should be more representative of the underlying stellar population. Indeed, this is a much greater concern for broad band colors of HSB disks, where the star formation rate and H II region covering factor are higher.

Due to bad weather, some galaxies were not observed through all filters. An effort was made to observe all galaxies over as long a spectral baseline as reasonable, so at least $B$ and $I$ were obtained. Table 3 lists the $B - I$ colors of those galaxies for which photometric data were obtained only in these bands. As before, nuclear, luminosity weighted, and area weighted colors are given.

## 4. STRUCTURE AND LUMINOSITY

The histograms of the structural parameters $(\mu_0, \alpha)$ for this sample are shown in Figure 3. The typical (median) values are $\mu_0 = 23.4$ mag arcsec$^{-2}$ and $\alpha = 2.5$ kpc. Making the same inclination correction (assuming disks are optically thin) Freeman (1970) and others have used, the median central surface brightness becomes $\mu_0^c = 23.8$ mag arcsec$^{-2}$. This is over $7\sigma$ from the "universal" value of $\mu_0^c = 21.65 \pm 0.3$ (Freeman 1970). It is also a full magnitude fainter than the average of the Romanishin et al. (1983) sample. That such remarkably LSB disks exist but have gone largely unrecognized highlights the importance of the selection effects which discriminate against stellar systems of low surface brightness (Disney 1976, Disney & Phillipps 1983).

Equally remarkable is that there is no preferred value of either $\mu_0$ or $\alpha$. There is a statistically insignificant peak in the $\mu_0$ histogram due to the sample selection. Objects with $\mu_0 < 23$ mag arcsec$^{-2}$ are rare by choice, while those with $\mu_0 > 24$ mag arcsec$^{-2}$ are obviously more difficult to identify and study in detail. Nonetheless, objects with central surface brightnesses as low as $\mu_0 \approx 24.5$ mag arcsec$^{-2}$ are present in the sample, and objects with $\mu_0 > 26$ mag arcsec$^{-2}$ are known to exist (e.g., Bothun, Impey, & Malin 1991).

Furthermore, the distribution of scale lengths is very broad. Smaller objects are more common than large ones (as expected from any reasonable intrinsic distribution), but both are present. It is a common misconception that LSB galaxies are dwarf galaxies, an unfortunate nomenclature which implies that all diffuse galaxies are intrinsically small. This is not the case, as the median scale length of 2.5 $h^{-1}$ kpc is indistinguishable from that of HSB spirals. We thus corroborate the result of Romanishin et al. (1983) that the sizes of LSB disks overlap those of HSB disks.

Moreover, there is no apparent correlation between the structural parameters $\alpha$ and $\mu_0$ (Figure 4), consistent with other studies (Impey, Bothun, & Malin 1988; Davies et al. 1988; Irwin et al. 1990; Bothun et al. 1991) which find that galaxies occupy *all* of observable parameter space. For comparison, the LSB sample of Romanishin et al. (1983) is also plotted in Figure 4, as is the predominantly HSB sample of van der Kruit (1987), with both adjusted to our adopted distance scale. There is, if anything, a tendency for the largest scale length disks to be low in surface density. This suggests that the dissipative formation of very extended, tenuous disks must have proceeded by gradual accretion without violent star formation or disruption by neighboring systems. This may reflect the need for large scale length disks to be isolated in order to collapse into a single object (see Bothun et al. 1990, Knezek 1992). More active systems would be subject to fragmentation into groups of smaller HSB galaxies rather than coalescence into enormous smooth structures like Malin 1 (Bothun et al. 1987), which would fall well below and to the right of the lower right corner of Figure 4 (Impey & Bothun 1989).

Despite being low in surface brightness, the integrated magnitudes of LSB galaxies can be quite bright (Figure 5). The large scale length disks have large total luminosities (and also tend to have fairly normal bulges, though this contribution to the total light is not included). At the least, it is clear that disk galaxies do exist over a very large range in each of scale length, absolute magnitude, and central surface brightness. Selection effects resulting from the brightness of the night sky strongly limit our knowledge of the physical properties and space density of objects over much of this range.

These considerations, combined with the apparent late collapse times of LSB galaxies (see below), lead us to suggest that the distribution of galaxies in the $(\mu_0, M_B)$ plane of Figure 5 is related to the initial conditions of galaxy formation. Very roughly, the luminosity corresponds to the total mass contained in the density perturbation from which the galaxy formed, and the surface brightness is related to the density contrast of that perturbation. This goes in the obvious sense that brighter galaxies are more massive, and LSB galaxies correspond to lower density contrasts at any given mass. The mass and the density contrast are unlikely to be entirely orthogonal in their effects, but neither are they parallel. This hypothesis is very successful in explaining the spatial distribution of LSB disks (Mo, McGaugh, & Bothun 1993; see also Bothun et al. 1993).

## 5. STELLAR CONTENT

LSB galaxies are very blue. Excluding the few systems (F568–6, UGC 5709, and UGC 6614) for which the bulge dominates the colors, the median colors are $B - V = 0.44$, $U - B = -0.17$, and $V - I = 0.89$. In comparison, an actively star forming Sc galaxy has typical colors of $B - V = 0.62$ and $U - B = 0$ (Huchra 1977), or $B - V = 0.50$ and $U - B = -0.2$ (Bothun 1982). The difference probably arises from the greater relative contribution of the bulge component within the photoelectric aperture of Huchra (1977). Regardless, a comparison Sc with active star formation is slightly redder on average than a more quiescent LSB disk of the same size. This is also true in $V - I$, the mean value for Sc galaxies being $V - I \approx 1$ (Han 1992). We have corrected the numbers of Han (1992) so that internal reddening is treated in the same fashion as for our galaxies (i.e., no correction), but no adjustment for differences in the treatment of galactic reddening (which is large for Han's sample) has been attempted. This would accentuate the difference between LSB and HSB Sc galaxies.

The blue colors of LSB disks are striking given their LSB nature and the paucity of H II regions relative to late type HSB spirals of comparable mass. The H$\alpha$ equivalent widths of LSB galaxies (McGaugh 1993) are lower than in the HSB galaxies in the sample of Kennicutt (1983) at similar color. The deviation of LSB galaxies from the color–H$\alpha$ equivalent width relation makes it is challenging to understand the stellar composition of LSB galaxies. It is likely that a variety of evolutionary histories are responsible given the broad distribution of colors in Figure 6. However, LSB disks as a group are quite blue, and we examine here the possible reasons for this.

### 5.1 Fading

The observed blue colors of LSB galaxies immediately rules out the most obvious possibility for their star formation history. That is, these are *not* HSB disks which have simply faded after the cessation of star formation. If this were the case, the galaxies would become progressively more red as their stellar populations evolve. For a normal stellar mass spectrum (IMF), $B - V$ will redden by $\sim 0.25$ for every 1 magnitude of fading in $\mu_B$. That there is no such trend of color with surface brightness in Figure 7a completely rules out this possibility.

While this is the case for the current sample, it does not mean that faded galaxies do not exist. Indeed, the pronounced lack of red LSB galaxies is one of the most curious aspect of the LSB surveys completed to date. Since these are performed on blue sensitive plates, it is quite conceivable that there is a color selection bias. For the median values of $\mu_0$, $\alpha$, and color observed here, the apparent diameter of a galaxy with the same size and bolometric luminosity would shrink by a factor of $\sim 2$ as it reddened by $\sim 1$ magnitude in $B - I$. Thus it would be visible in only one eighth of the volume that the blue galaxies could be detected over in diameter limited catalogs of the sort employed here, and should be correspondingly underrepresented.

While there are indeed few red galaxies in our sample, there is not the sharp fall off in numbers with redder color expected by this reasoning. However, the shrinkage of apparent diameters depends sensitively on both the bolometric surface brightness and the unknown details of the spectral energy distributions of the hypothesized red galaxies, and is therefore difficult to calculate. Our sample is small and not complete anyway, so such an effect can not be ruled out. If important, our current sample may just represent the blue extreme of a much larger variety of as yet undiscovered LSB galaxies.

The small volume over which red LSB galaxies can be detected implies a potentially large space density of such objects. This is largely unconstrained by current observations, and could be related to the excess population of faint blue galaxies (e.g., Tyson 1988; Lilly, Cowie, & Gardner 1991; Metcalfe et al. 1991; Colless et al. 1991) observed at intermediate redshifts. These might very well fade in surface brightness and redden by the present epoch. On the other hand, the density of LSB disks of all colors seem likely to have been underestimated. Since the colors and luminosities of the blue sample presented here are similar to those of the excess population (Lilly 1993), there may be a direct connection without invoking large amounts of evolution. The star formation history inferred for LSB galaxies (§5.5) suggests that they would remain blue out to intermediate redshift.

Note also that for the present sample color is not correlated with size as measured by the exponential scale length $\alpha$ (Figure 7b). The two giant disks, F568–6 (also known as Malin 2; Bothun et al. 1990) and UGC 6614 are both red, but giant disks tend to have steep color gradients (Bothun et al. 1990, Knezek 1992, Sprayberry et al. 1993) indicating a low degree of mixing of their stellar populations. The lack of correlations between the structural parameters and color indicates that a galaxy's evolutionary history is not uniquely determined by its size, and that color–luminosity relations should be reconsidered with due consideration for the sample selection effects involved.

### 5.2 *The IMF*

One possible reason for the blue colors of LSB galaxies is that their IMF is somehow different. It has been suggested that the low surface brightnesses of these galaxies could be due to a dearth of high mass stars (Romanishin et al. 1983, Schombert et al. 1990). However, matching the blue colors would require an IMF biased in *favor* of high mass stars. No variation in the IMF is inferred from the H II region observations (McGaugh 1993), which indicate that high mass stars are present. If anything, the emission properties of the H II regions would require an excess of high mass stars, while the low observed metallicities would suggest a lack of them. Neither of these are consistent with the mass to light ratios, which are roughly normal (McGaugh 1992).

Arimoto & Tarrab (1990) have made a model which specifically addresses this quixotic situation, wherein metal production is suppressed until recent epochs by an IMF which flattens with time. Some choices of the parameters of this model adequately match the $B - V$ and $U - B$ colors of LSB galaxies (though mostly they are too blue), but $V - I$ is always predicted to be too red. It is in principle possible to match nearly any observation if sufficient violence is done to the IMF (von Hippel & Bothun 1990), but the requirements in this case are conflicting. Since there is no convincing evidence for variation in the IMF of LSB (McGaugh 1993), or indeed, *any* galaxies (McGaugh 1991), it is highly unlikely that this be the major cause of the observed blue colors.

### 5.3 *Peculiar Phases of Stellar Evolution*

Schombert et al. (1990) explored the possibility that the blue colors of LSB galaxies could be caused by an excess of blue horizontal branch (BHB) stars. This could be brought about by the low metallicities of LSB galaxies (such stars being important contributors to the total light in metal poor globular clusters) or some other peculiarity. Without direct knowledge of the distribution of stars in the HR diagram in LSB galaxies, it is difficult to discount the possibility that there exists a population such as BHB stars which strongly affects the colors.

In this sense, the problem is analogous to that of the ultraviolet excess in elliptical galaxies. Some stage of stellar evolution (presumably in the later, less understood periods) which is somehow emphasized in the galaxies in question is required to noticeably alter the spectral energy distribution expected from all other stages. If BHB stars are for some reason more common in LSB galaxies, this might result in the observed colors.

This possibility could be tested by space based ultraviolet imaging of LSB galaxies. For the present, we note that the ultraviolet excess in ellipticals is apparently produced by a small population of extreme BHB stars which are probably metal *rich* (Ferguson & Davidsen 1993; see also Bertelli, Chiosi, & Bertola 1989). Though this is opposite the sense expected for LSB galaxies, the idea that there is something peculiar about some of the stars in galaxies which are themselves far from "normal" cannot be ruled out. It seems unlikely, however, that an abnormal post main sequence phase could have such a strong effect on the optical colors (see Appendix), and whatever peculiar population is invoked (subdwarf B stars?) must be widely enough spaced to meet surface brightness and mass to light ratio constraints.

### 5.4 *Metallicity*

The effects of metallicity and age on the color of the composite, unresolved populations in galaxies are difficult to disentangle. Colors are sensitive to both the mean age and the shape of the age distribution, with blue colors generally indicating young ages. Low metallicity can also cause blue colors, through changes in line blanketing and the positions of both the main sequence (Maeder 1990) and the giant branch (Bothun et al. 1984).

Nonetheless, $B - V$ should be more sensitive to mean age, while $V - I$ will be sensitive to both the position of the giant branch (metallicity) and its degree of development (age). $U - B$ will be sensitive to both metallicity and the presence of hot, very young stars. Fortunately, we are not restricted to the broad band colors alone and also know that the metallicity in the gas, and presumably in the stars, is low (McGaugh 1993).

The lowest metallicity globular clusters have integrated colors of roughly $B - V \approx 0.65$, $U - B \approx 0.1$, and $V - I \approx 0.95$ (Reed 1985). This is close to $V - I$ in LSB galaxies, but rather red in the other colors. This suggests that LSB galaxies could possess an old population similar to that of the metal poor globular clusters which dominates $V - I$, but that $B - V$ and $U - B$ have contributions from more recent star formation. While LSB galaxies are not as metal poor as the bluest globular clusters, it is reasonable to suppose this to be the case at some level.

The question becomes the degree to which age and metallicity contribute. Since we have information about the metallicity (McGaugh 1993), we can directly measure its effect. There is no correlation between metallicity (as traced by the oxygen abundance in the H II regions) and any of the colors, including $V - I$ (Figure 8). Hence, while metallicity must be a contributing factor, it cannot be the dominant cause of the blue colors.

### 5.5 *Star Formation History*

A young mean age is thus left as the primary cause of the blue colors in LSB galaxies. This can take the form of a burst of star formation imposed (or not) on an older population, or a more continuous star formation history with most of the light (if not necessarily the mass) being contributed by the younger generations. A series of sharp bursts is evidently not responsible, since the LSB nature of the disks implies that they should be far removed (and hence faded red) from any preceding burst. Thus a more continuous star formation rate is indicated.

While an old, metal poor population is likely to exist in most of these galaxies, the complex star formation histories do seem to be weighted towards youth. That $U - B < 0$ in most cases immediately indicates that much of the light is being produced by stars that have formed relatively recently, and that the star formation rate in the last $10^8$ yr is comparable to or larger than that averaged over a Hubble time (Larson & Tinsley 1978). The median colors are remarkably blue for composite stellar populations, and *some* LSB galaxies are *extremely* blue. It follows that these may have a very young mean age, as was recently argued to be the case for the H I cloud in Virgo (Salzer et al. 1991), a typical LSB galaxy by the standards of this sample.

Note that these blue colors occur despite the low covering factor of bright knots of OB associations relative to that in Sc galaxies. The colors are not skewed by these HSB knots of very short lived stars, and must be representative of the entire population. Since the emission equivalent widths of H II regions in LSB galaxies are on average lower than those in HSB galaxies (McGaugh 1993; cf. Phillipps & Disney 1985), and the numbers of H II regions are lower (see images in McGaugh, Bothun, & Schombert 1993), LSB galaxies have lower star formation rates per unit mass than HSB galaxies of similar color. Thus the general disk population of LSB galaxies must be producing their very blue colors, so the stellar population as a whole must have a young mean age. This is consistent with the morphology of these galaxies, which does not change much between the $U$ and $I$ filters. This indicates fairly homogeneous populations which lack the conspicuous red old disk population typically seen in HSB galaxies. Indeed, the total mass contained in any old, dim population is rather tightly constrained by the modest mass to light ratios (McGaugh 1992; cf. Salzer et al. 1991).

A simple star formation history which approximately matches the observed properties of LSB galaxies is one with a low, constant star formation rate (SFR). Larson & Tinsley (1978) considered models covering the range of star formation histories from single bursts to continuous star formation at a constant rate. Two of their models with constant SFRs can provide an adequate match to the LSB data. Their 5 Gyr model has $B - V = 0.45$, $U - B = -0.16$, and $M/L_B \sim 1$, while their 10 Gyr model has $B - V = 0.50$, $U - B = -0.09$, and $M/L_B \sim 2$ (using their empirical values). The 5 Gyr model is somewhat better, providing an almost exact match to the median LSB colors. We can use the $M/L$ ratios to determine the absolute star formation rate required to match the observed luminosity and surface brightness of LSB disks. A disk galaxy with scale length $\alpha = 2.5$ kpc and SFR $\psi = 0.2 \, M_\odot \mathrm{yr}^{-1}$ will have a luminosity $L_B \approx 10^9 L_\odot$ and central surface brightness $\mu_0 \approx 23.4$ in both the 5 and 10 Gyr cases (since the difference in $M/L$ compensates for the lesser amount of mass converted into stars after 5 Gyr). These numbers are quite reasonable for this sample, so it seems that LSB galaxies have been forming stars at a low, approximately constant rate for somewhat less than a Hubble time.

While the $B - V$ and $U - B$ colors indicate that the present rate of star formation is not significantly less than that in the past, they do not much constrain the detailed shape of the star formation history. But we also have $I$, and all of the LSB galaxies are bluer in $B - I$ than the HSB disks (like our own galaxy) which have been forming stars at a constant rate for long enough to build up a significant old red disk. Given that the 5 Gyr model gives a better fit than the 10 Gyr model, and that metallicity is unable to explain all of the difference in the colors, it seems likely that LSB galaxies on average commenced their star formation later then HSB disks. Late formation naturally explains the low metallicities of LSB systems, as well as their low surface densities (McGaugh et al. 1993) and spatial distribution (Mo et al. 1993).

Figure 9 illustrates simple star formation histories in which the SFR is a constant, but in which one galaxy (A) started to form stars at an earlier epoch than another (B). As long as the stellar physics and the IMF are not too dissimilar, the younger population will be bluer by an amount determined by the difference in turn on times (see Appendix). Physically, this stems from the underpopulation of the giant branch relative to the main sequence, which decreases the flux in $I$ relative to that in $B$. This relative dearth of giants occurs because the number of post main sequence stars is determined by the rate at which stars leave the main sequence, which in the case of continuous star formation is proportional to the range of turn off masses present in the composite population. This in turn is directly determined by the spread in ages, which is obviously less in any population which commences star formation later. From the present data it is impossible to distinguish star formation history B from some more complicated form like C in Figure 9, but the bulk of the star formation does seem to be weighted towards recent epochs. Such a contrast in the star formation rate between the first and second half of the Hubble time is reminiscent of that determined for the LMC by direct observation of the distribution of stars in the HR diagram (Bertelli et al. 1992).

More sophisticated population synthesis models are clearly desirable. Those currently available generally assume a star formation rate which declines with time as gas is consumed: $\psi \propto M_{gas}^n$. While this seems a very reasonable assumption, it turns out to be a very bad one for LSB galaxies (van der Hulst et al. 1993; McGaugh et al. 1993) and disks in general (van der Hulst et al. 1987, Kennicutt 1989). It would be useful if future modeling explored other forms for the star formation history, and a broad range of metallicities. Interestingly, the model which best matches the very blue LSB galaxies is that of Katz (1992). This is for disks early in their evolution (corresponding to HSB spirals at $z \gtrsim 1$), consistent with a late turn on time for LSB galaxies.

## 6. CONCLUSIONS

We have presented broad band observations of a sample of low surface brightness galaxies. We find that these objects are reasonably well described as exponential disks, and that often their surface brightness profiles are purely exponential without significant bulge components. These galaxies have no preferred value of either scale length or central surface brightness. The median scale length ($2.5\,h^{-1}$ kpc) is quite large and indistinguishable from that of "normal" high surface brightness disks (i.e., these are *not* dwarf galaxies), despite the median central surface brightness ($\mu_0$ = 23.4 mag arcsec$^{-2}$) being $\gtrsim 7\sigma$ below the canonical Freeman (1970) value.

There are no correlations between the structural parameters $\mu_0$ and $\alpha$, indicating that disk galaxies exist over a large range of both size and stellar surface density. That LSB galaxies cover such a wide range in these parameters indicates that there are many evolutionary paths which do not lead to the Hubble Sequence. The striking physical properties of LSB disks also make it clear that the selection effects imposed by the arbitrary level of the night sky brightness severely limit our knowledge of the full range of galaxy properties. Probing beneath the sky leads us to suggest that the luminosity and surface brightness of disks are related to the initial conditions of galaxy formation, mass and collapse time, respectively.

LSB galaxies are observed to be very blue, with median colors $B - V = 0.44$, $U - B = -0.17$, and $V - I = 0.89$. The spread in colors is large, indicating generally complex star formation histories. Color is not related to surface brightness, ruling out fading scenarios. Variation in the IMF or stellar evolution (leading, for example, to an overdeveloped BHB) is unlikely to be the cause of the observed blue colors. The low metallicities of LSB galaxies must contribute to some extent to their blue colors, but cannot be the primary cause as color and metallicity are uncorrelated. This leaves a young mean age as the primary cause. The most plausible scenario is that LSB galaxies form stars at a constant rate after a late collapse, and/or that the star formation rate has been gradually increasing with time.

We thank Jim Schombert for being a dude. We also thank Peter Mack and Bob Barr for the ever improving situation at MDM Observatory, which facilitated many productive visits. This research was supported in part by NSF grant AST 90–05115. It has made use of the NASA/IPAC Extragalactic Database (NED) which is operated by the Jet Propulsion Laboratory, California Institute of Technology, under contract with the National Aeronautics and Space Administration.

# APPENDIX

Here we apply the formalism of analytic population synthesis (Tinsley 1972; Tinsley & Gunn 1976) to the case of constant star formation. This is useful for illustrating the effects of variable turn on times on the relative balance between the contributions of main sequence (MS) and post main sequence (PMS) stars, and the subsequent impact on the colors of a stellar system composed of many generations.

It is standard and convenient to adopt power law relations which describe the stellar population and stellar physics. These are of course approximations, and single slope power laws do not provide perfect fits for the entire range of stellar properties, which may also have other dependencies than those primarily assumed. This is, however, quite adequate for our current purposes.

We shall decompose the stellar birthrate function, which describes the stellar population, into a mass spectrum (the IMF) and a star formation rate (SFR). The IMF gives the relative number of stars of different masses,

$$\frac{dN}{dm} = \phi(m) = \phi_0 m^{-(1+x)}, \tag{A1}$$

which is assumed to be valid between some limiting upper ($m_u$) and lower ($m_\ell$) mass. The precise values of $x$ and $m_u$ are controversial, even though $m_u$ may only be determined in a probabilistic way from the distribution defined by $x$. The precise values of the parameters are not important to what follows, though we will assume that $x$ is constant and has an astrophysically plausible value (i.e., low mass stars are more frequent than high mass stars, so $x > 0$). We consider SFRs of the form illustrated in Figure 9,

$$\psi_j(t) = \psi_0 u(t_j), \tag{A2}$$

where $u(t_j)$ is the step function. Thus star formation model $j$ is characterized by the turn on time $t_j$.

The stellar physics is encapsulated in the main sequence mass–luminosity and mass–lifetime relations. Respectively, these are

$$\ell(m) = \ell_0 m^{\xi_k} \tag{A3}$$

and

$$m(t) = \mathcal{M}_0 t^{-\theta}. \tag{A4}$$

Note that we allow the slope of the mass–luminosity relation to depend on the observational bandpass $k$. We will assume that the luminosity of post main sequence stars is adequately modeled by a constant luminosity $\mathcal{L}^{\rm PMS}$ for each post main sequence phase over the (relatively short) lifetime of that phase $\tau^{\rm PMS}$.

Combining these allows us to write a simple expression for the luminosity produced by a single generation of stars. For the main sequence, this is simply the convolution of the IMF and mass–luminosity relation:

$$L_i^{\rm MS} = \int_{m_\ell}^{m_\tau} \phi(m)\ell(m)dm. \tag{A5}$$

The upper limit is set by the main sequence turn off mass $m_\tau$, so the value of $m_u$ is irrelevant. Similarly, since we are interested in colors and not mass to light ratios, we can set $m_\ell \to 0$, as low mass stars may represent much mass but produce very little light. Substituting into equation (A5) the relations defined above, we have

$$L_i^{\rm MS} = \phi_0 \ell_0 \int_0^{m_\tau} m^{\xi_k - 1 - x} dm, \tag{A6}$$

which gives

$$L_i^{\rm MS} = \frac{\phi_0 \ell_0}{\xi_k - x} m_\tau^{\xi_k - x}. \tag{A7}$$

To get the contribution from all generations, we must convolve equation (A7) with the weighted contribution of each, so

$$L_{\rm (total)}^{\rm MS} = \sum w_i L_i = \int w(t) L(t) dt. \tag{A8}$$

Here, $L$ depends on $t$ through equation (A4), and the weights are just the normalized SFRs $w_j(t) = \psi_j(t)/<\psi_j>$. This gives

$$L^{\rm MS} = \frac{\phi_0 \ell_0}{<\psi_j>(\xi_k - x)} \int_0^{t_0} \psi(t) m_\tau(t)^{\xi_k - x} dt, \tag{A9}$$

where $t_0$ is the present epoch. Replacing $m_\tau$ with equation (A4) and allowing the chosen step function form of the SFR (equation A2) to operate on the limits of integration,

$$L_{jk}^{\rm MS} = \frac{\phi_0 \ell_0 \psi_0 \mathcal{M}_0^{\xi_k - x}}{<\psi_j>(\xi_k - x)} \int_{t_j}^{t_0} t^{-\theta(\xi_k - x)} dt. \tag{A10}$$

Thus we have an expression for the main sequence $k$-band luminosity of a composite stellar population specified by star formation history $j$:

$$L_{jk}^{\mathrm{MS}} = \frac{\phi_0 \ell_0 \psi_0 \mathcal{M}_0^{\xi_k - x}}{<\psi_j>(\xi_k - x)(\theta x - \theta \xi_k + 1)} t^{\theta x - \theta \xi_k + 1} \bigg|_{t_j}^{t_0}. \qquad (A11)$$

Note that the limits do matter as $t_j$ specifies the star formation history.

The numbers of post main sequence stars from any generation follows directly from the rate at which stars are turning off the main sequence and the amount of time they spend in each PMS phase, $\tau^{\mathrm{PMS}}$:

$$N_i^{\mathrm{PMS}} = \frac{dN}{dm}\bigg|\frac{dm}{dt}\bigg|_{m_\tau} \tau^{\mathrm{PMS}} = -\phi(m_\tau)\frac{dm_\tau}{dt}\tau^{\mathrm{PMS}}. \qquad (A12)$$

Substituting the appropriate expressions, this becomes

$$N_i^{\mathrm{PMS}} = \phi_0 \tau^{\mathrm{PMS}} \mathcal{M}_0^{-x} \theta t^{\theta x - 1}, \qquad (A13)$$

which when multiplied by the average luminosity of each phase gives the total luminosity produced by that branch,

$$L_i^{\mathrm{PMS}} = \phi_0 \tau^{\mathrm{PMS}} \mathcal{L}^{\mathrm{PMS}} \mathcal{M}_0^{-x} \theta t^{\theta x - 1}. \qquad (A14)$$

This can be weighted by the star formation history as well,

$$L_{jk}^{\mathrm{PMS}} = \frac{\psi_0}{<\psi_j>}\phi_0 \tau^{\mathrm{PMS}} \mathcal{L}^{\mathrm{PMS}} \mathcal{M}_0^{-x} \theta \int_{t_j}^{t_0} t^{\theta x - 1} dt, \qquad (A15)$$

which gives simply

$$L_{jk}^{\mathrm{PMS}} = \frac{\psi_0}{<\psi_j>}\frac{\phi_0}{x\mathcal{M}_0^x} \tau^{\mathrm{PMS}} \mathcal{L}^{\mathrm{PMS}} t^{\theta x} \bigg|_{t_j}^{t_0}. \qquad (A16)$$

The total luminosity contributed by each post main sequence phase is about equally divided between the giant branch, the HB, and the AGB (Renzini & Buzzoni 1986). As long as this is true (see §5.3), the light produced by PMS stars will always be redder than that from the MS since the "average" PMS star is closer to the giant branch than the horizontal branch (which itself is often red). Dividing equation (A16) by equation (A11) gives the ratio of PMS to MS luminosity:

$$\frac{L_{jk}^{\mathrm{PMS}}}{L_{jk}^{\mathrm{MS}}} = \frac{\tau^{\mathrm{PMS}} \mathcal{L}^{\mathrm{PMS}}}{\mathcal{M}_0 \ell_0} \frac{(\xi_k - x)(\theta x - \theta \xi_k + 1)}{x} \frac{t^{\theta x}\big|_{t_j}^{t_0}}{t^{\theta x - \theta \xi_k + 1}\big|_{t_j}^{t_0}}. \qquad (A17)$$

This gives the relative weight of the predominantly red and blue components, and hence some indication of the composite color. Assuming that the IMF is the same, and that differences in metallicity do not excessively alter the composite spectral energy distribution of the various phases, we can cancel all but the temporal terms, and compare the effects of different star formation histories on different filters. For the cases illustrated in Figure 9, $j = \{A, B\}$, $t_B > t_A$, and filters $k = \{B, I\}$,

$$\left. \frac{\frac{L^{\mathrm{PMS}}}{L^{\mathrm{MS}}}|_A}{\frac{L^{\mathrm{PMS}}}{L^{\mathrm{MS}}}|_B} \right|_B < \left. \frac{\frac{L^{\mathrm{PMS}}}{L^{\mathrm{MS}}}|_A}{\frac{L^{\mathrm{PMS}}}{L^{\mathrm{MS}}}|_B} \right|_I \tag{A18}$$

as long as $\xi_B > \xi_I$, which is guaranteed by the main sequence temperature–luminosity relation. Note that differences in the absolute rates of star formation are irrelevant, since $\psi_0$ has already been canceled out of equation (A17) — only the shape of the star formation history matters. Equation (A18) simply states that the older population (A) will produce relatively more red light. This arises from the increasing contribution of (red) post main sequence stars to the total luminosity as time goes on, i.e., the increasingly populace giant branch (cf. Renzini & Buzzoni 1986). Hence the blue colors and lack of diffuse red disks in LSB galaxies can result from late collapse and ignition of star formation.

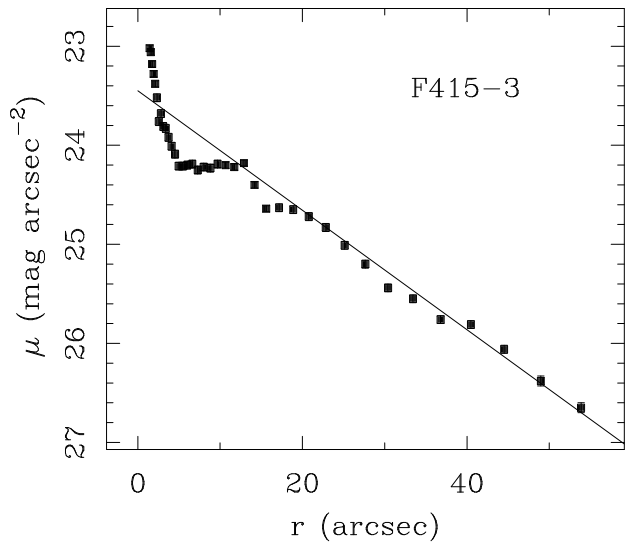
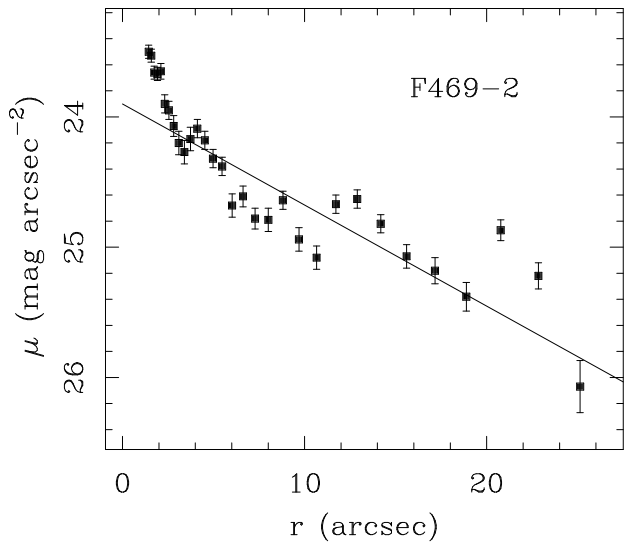
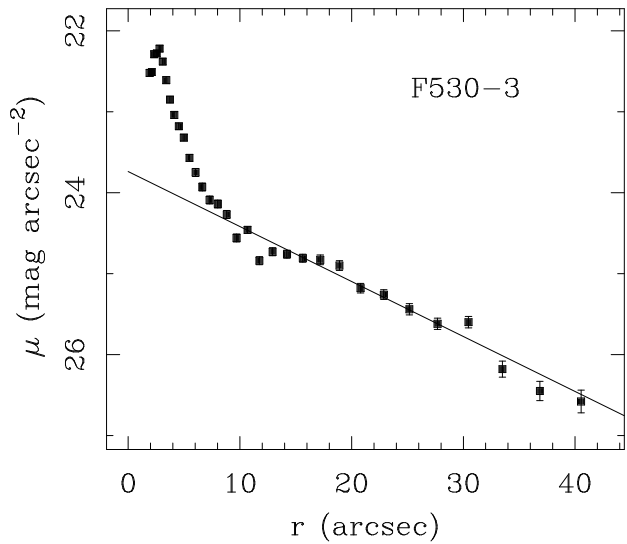
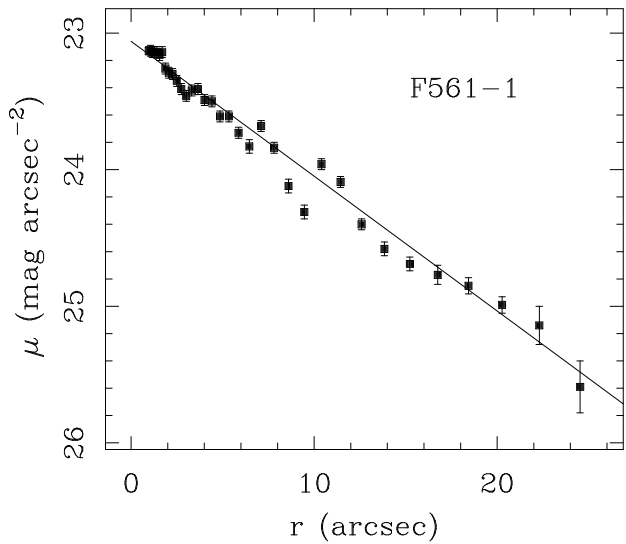
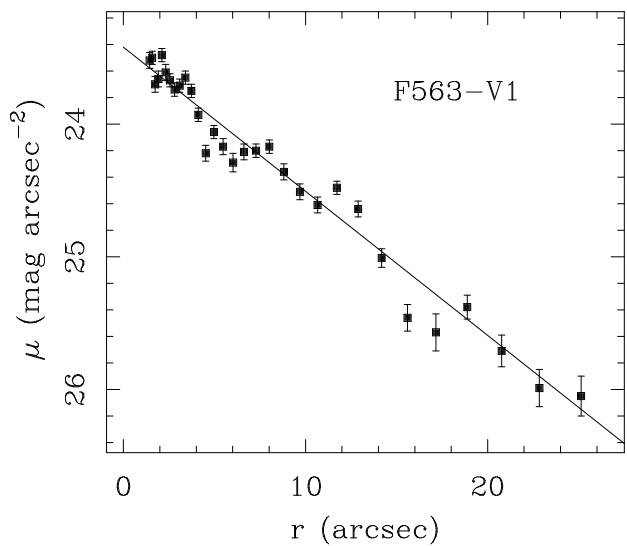
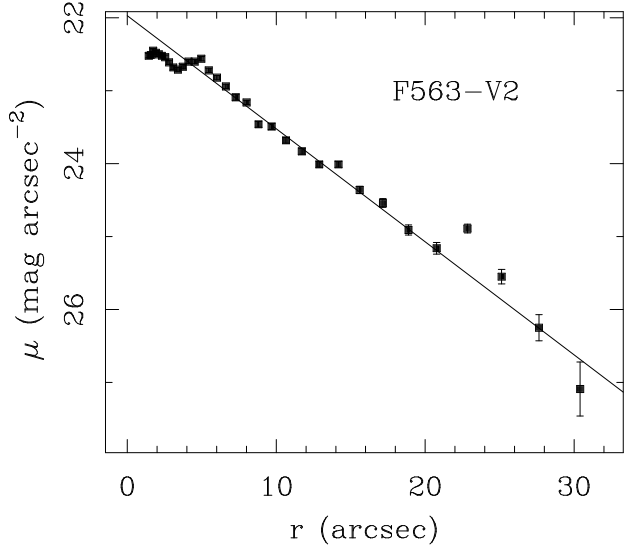

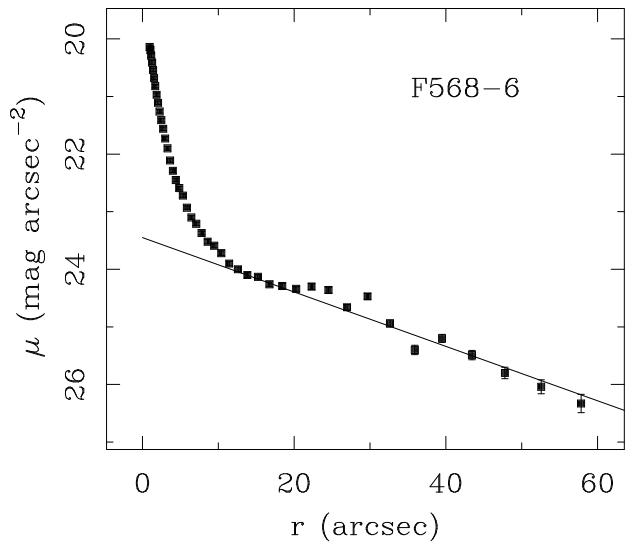
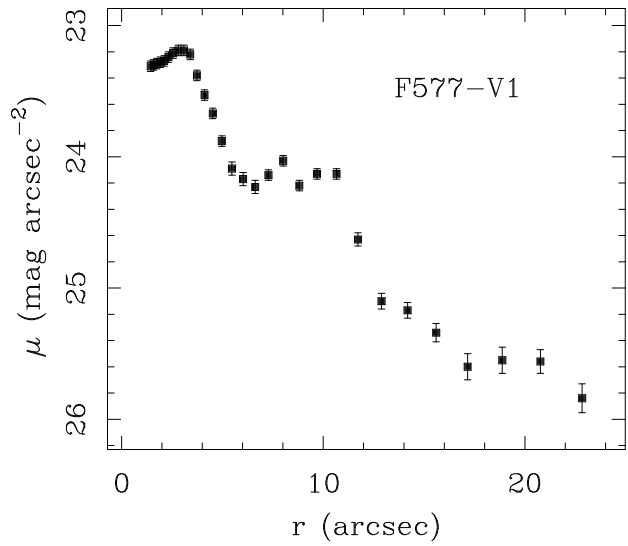
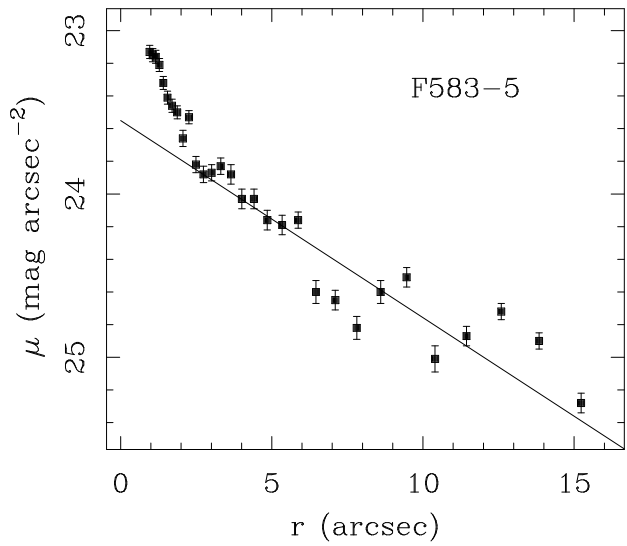
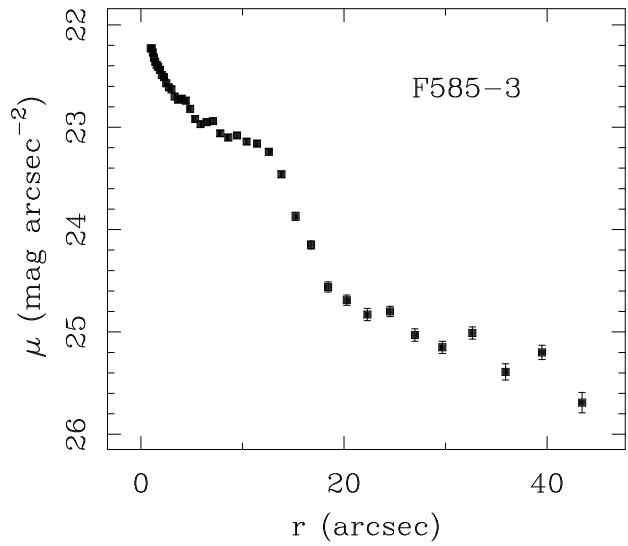
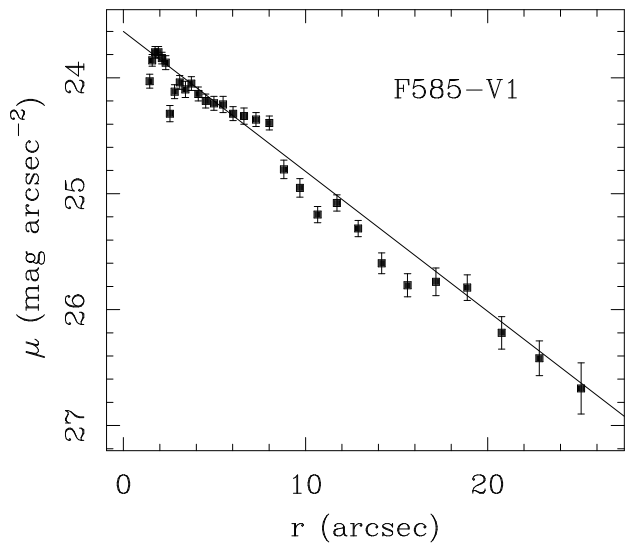
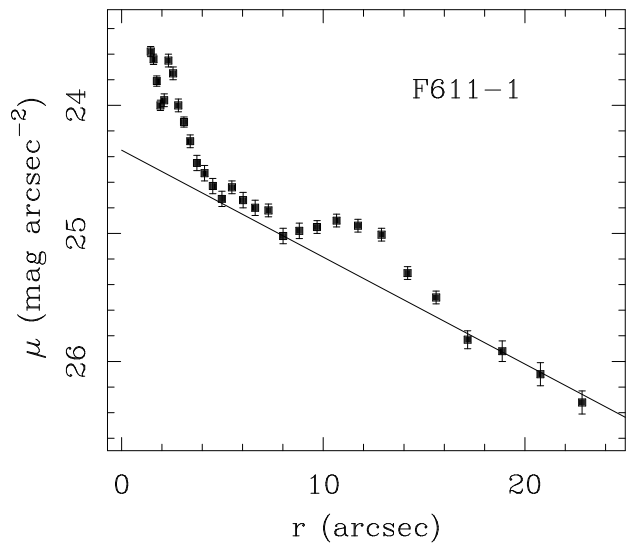

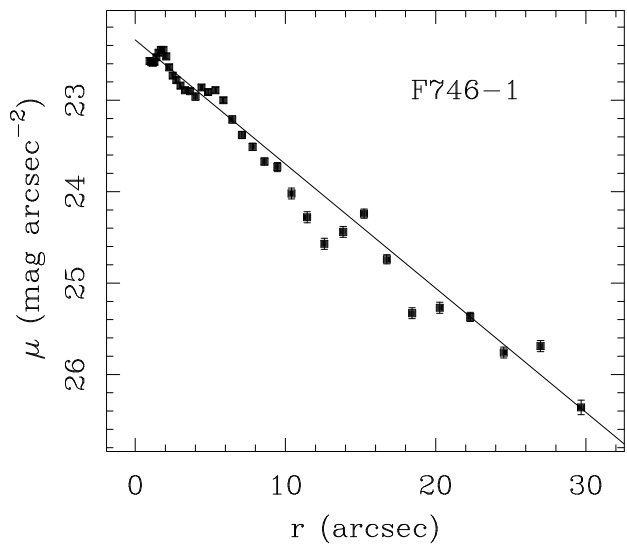
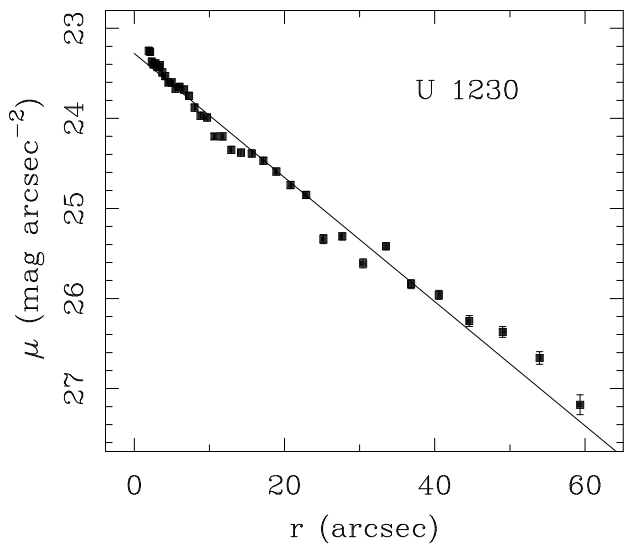
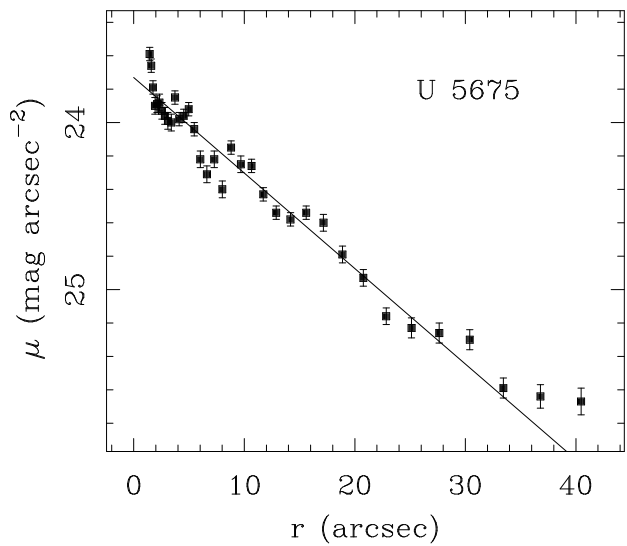
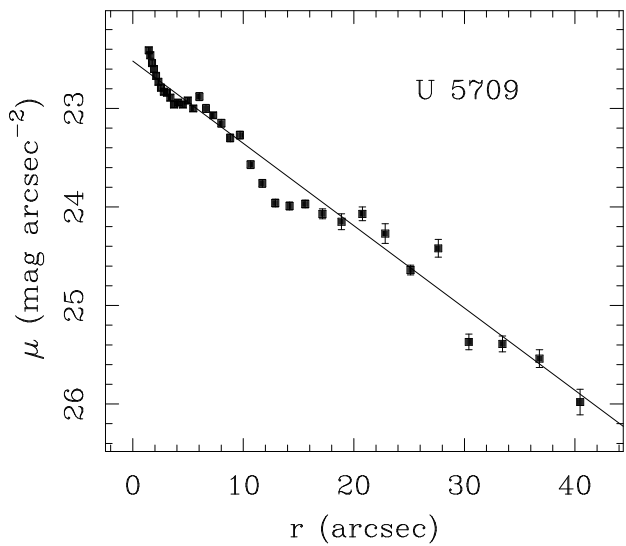
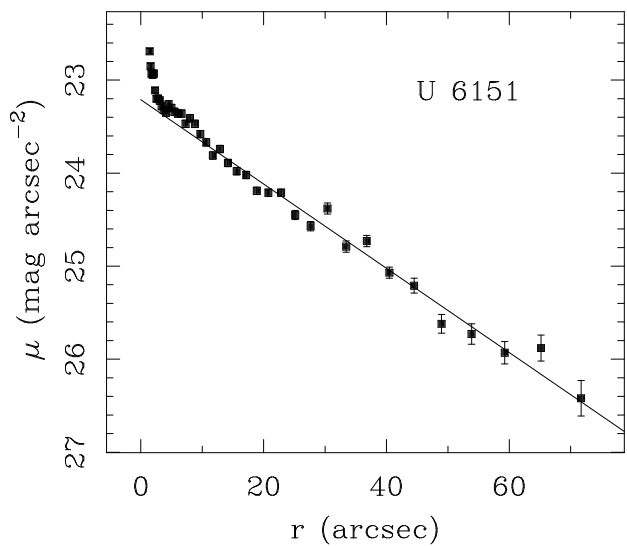
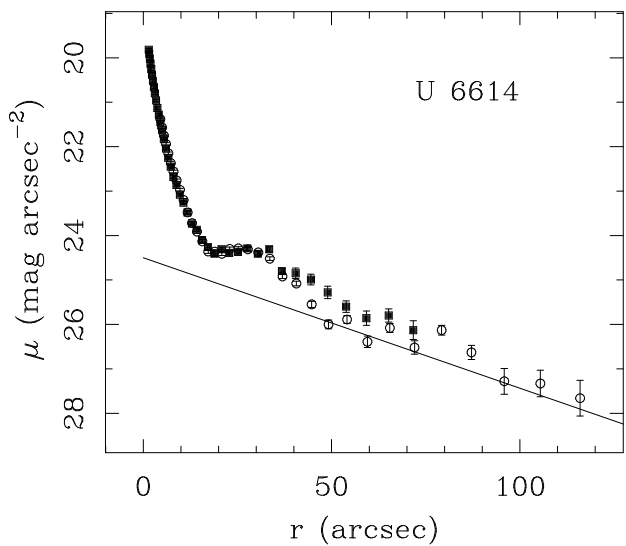

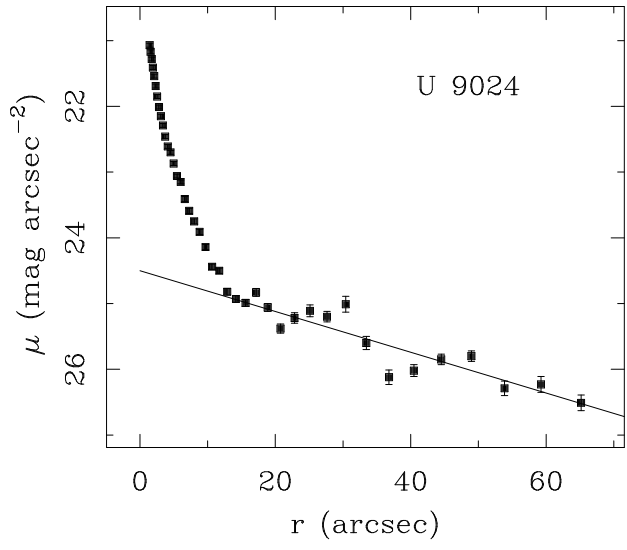
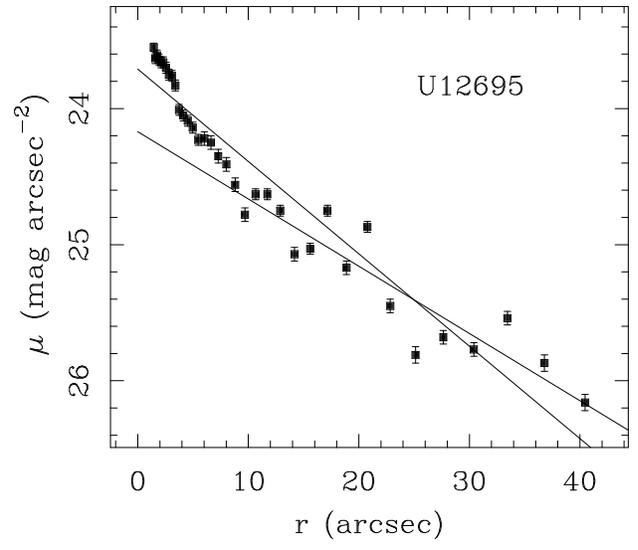

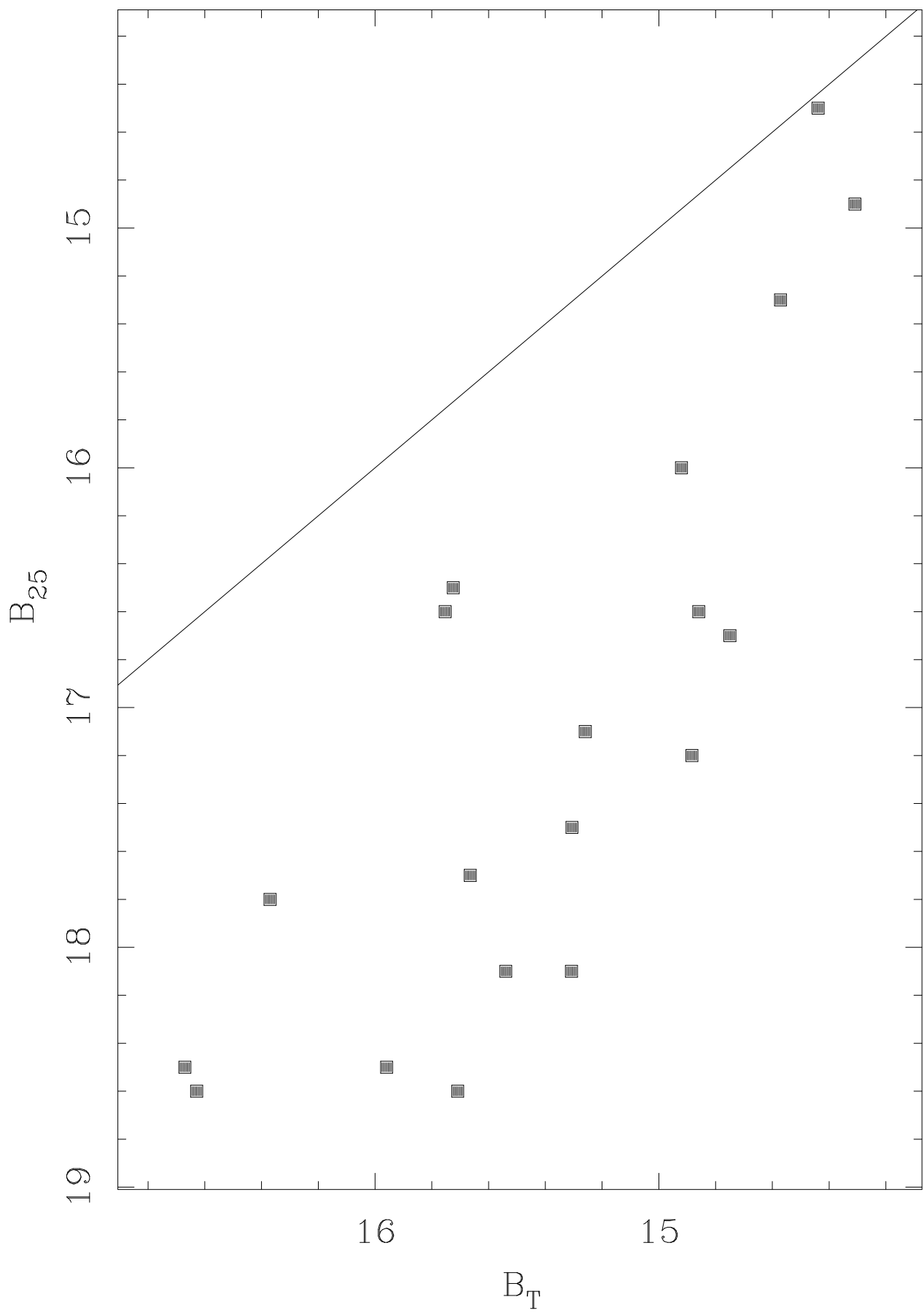

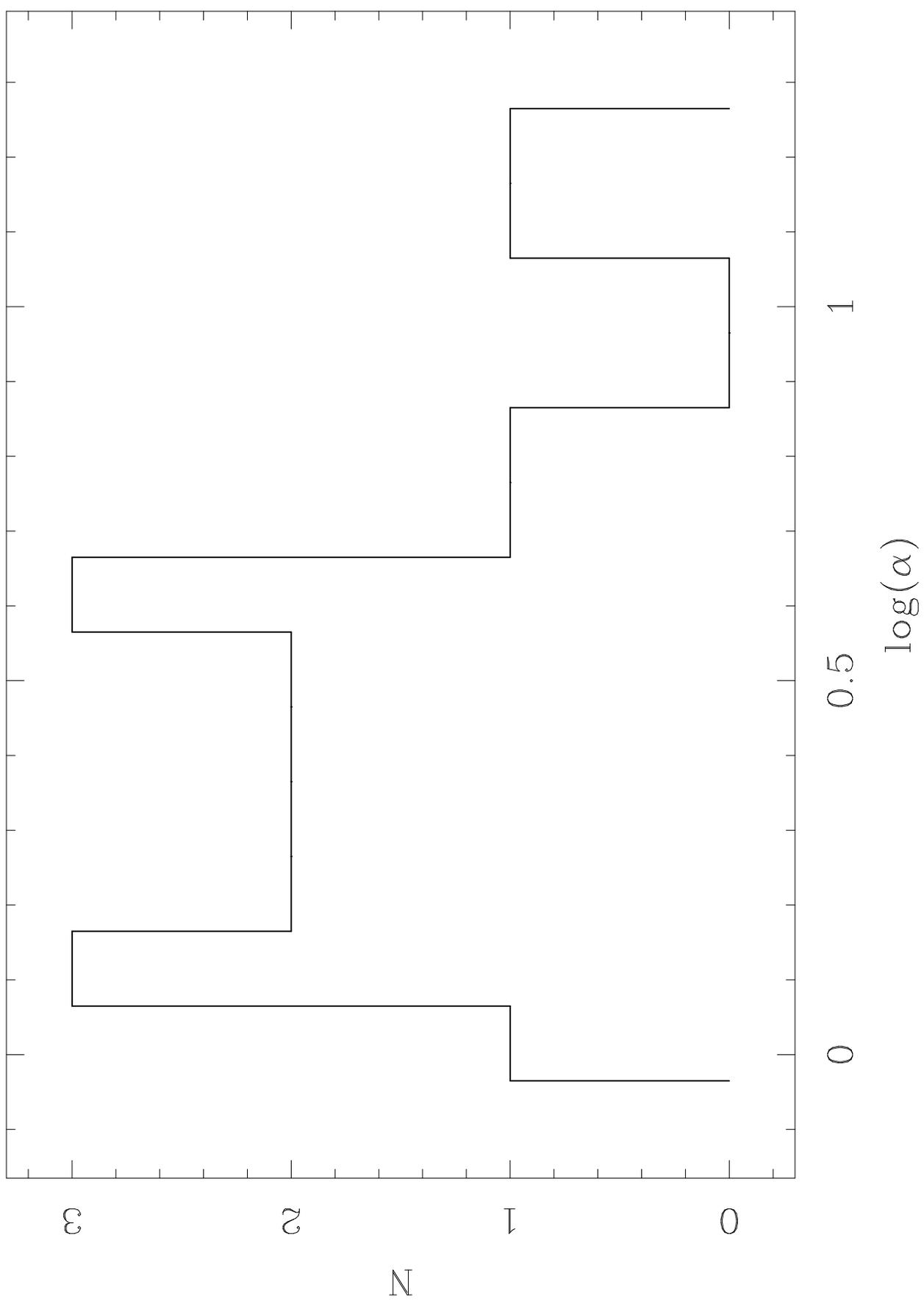

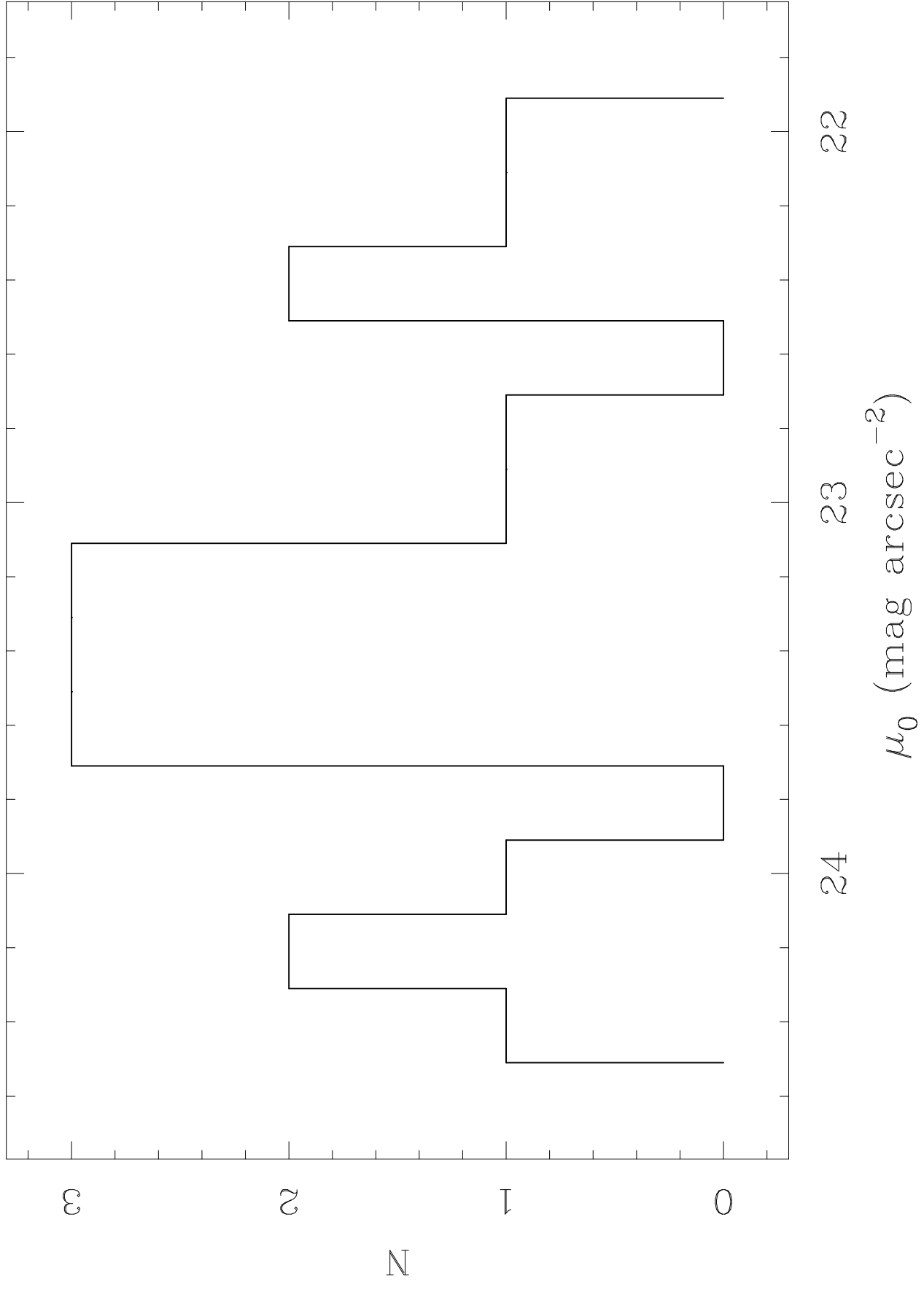

$\mu_0$ (mag arcsec$^{-2}$) vs $\log(\alpha)$

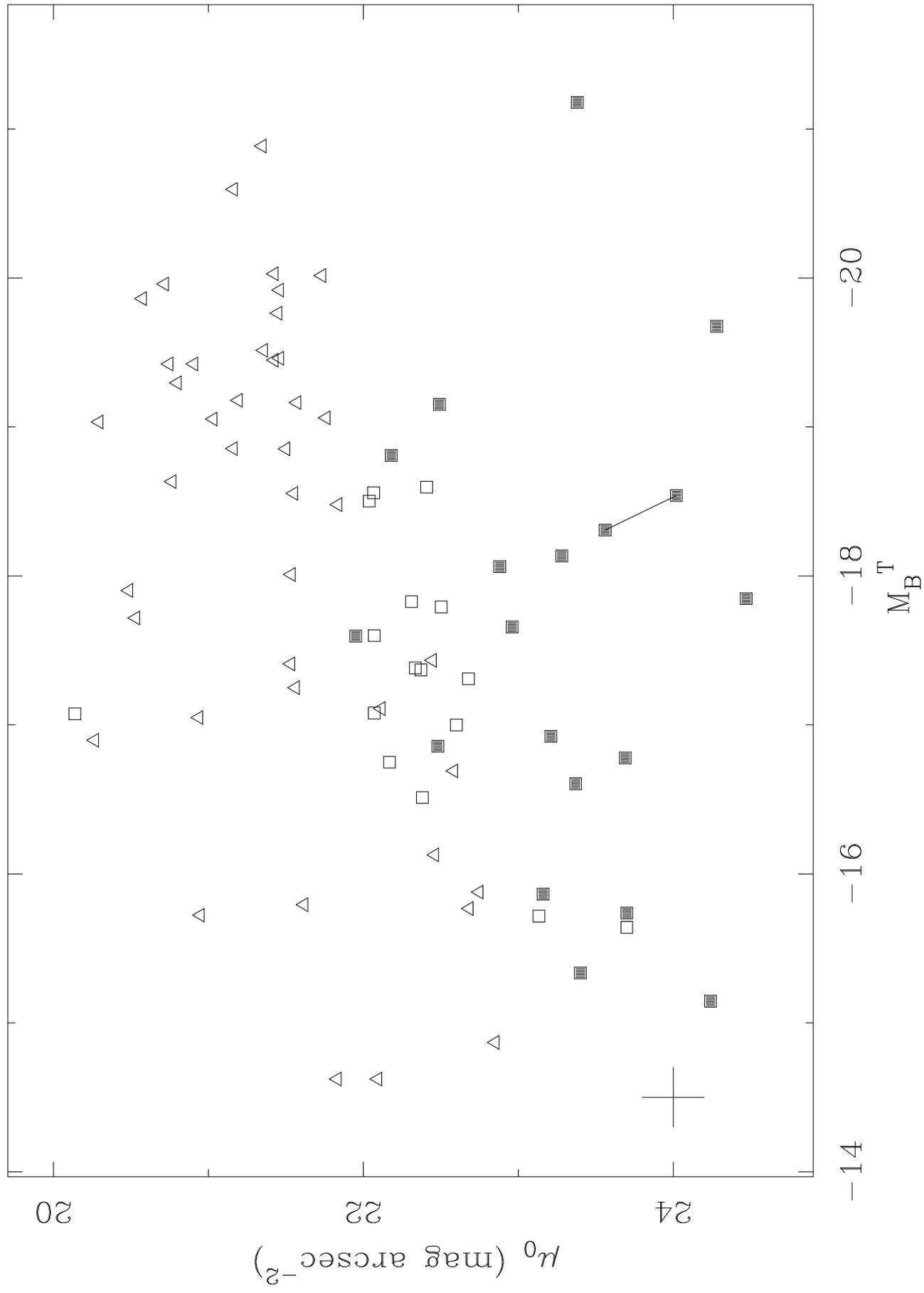

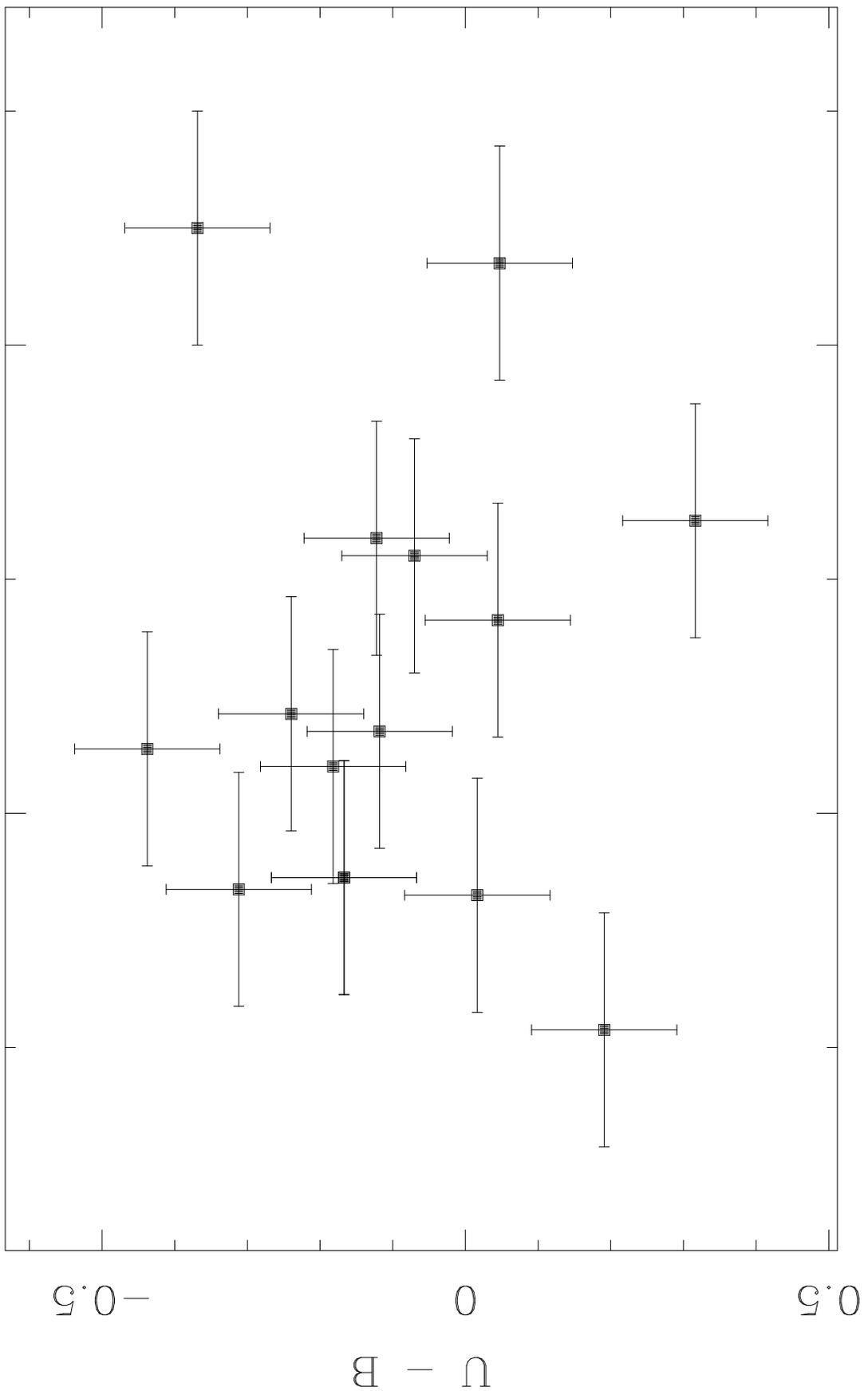

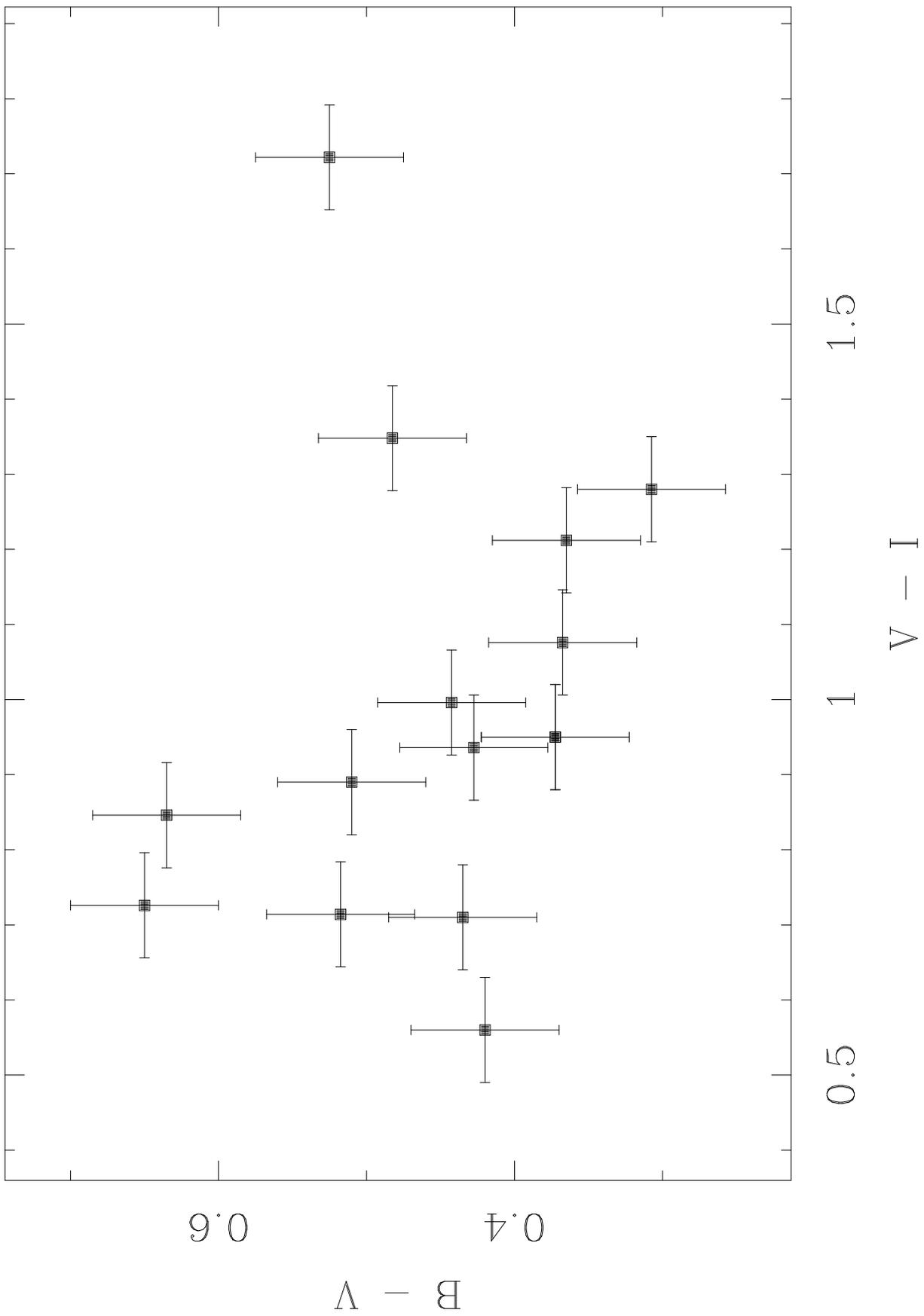

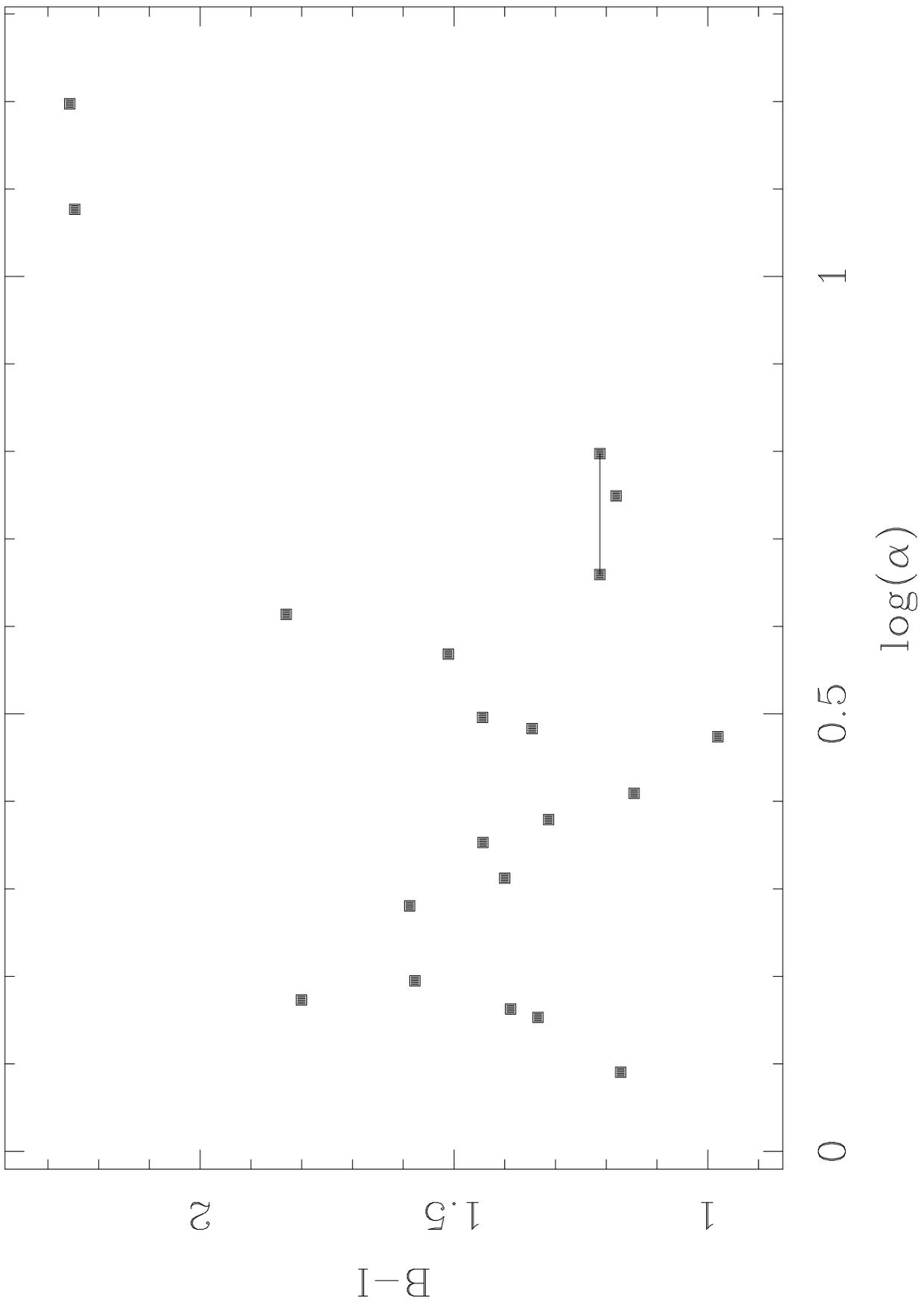

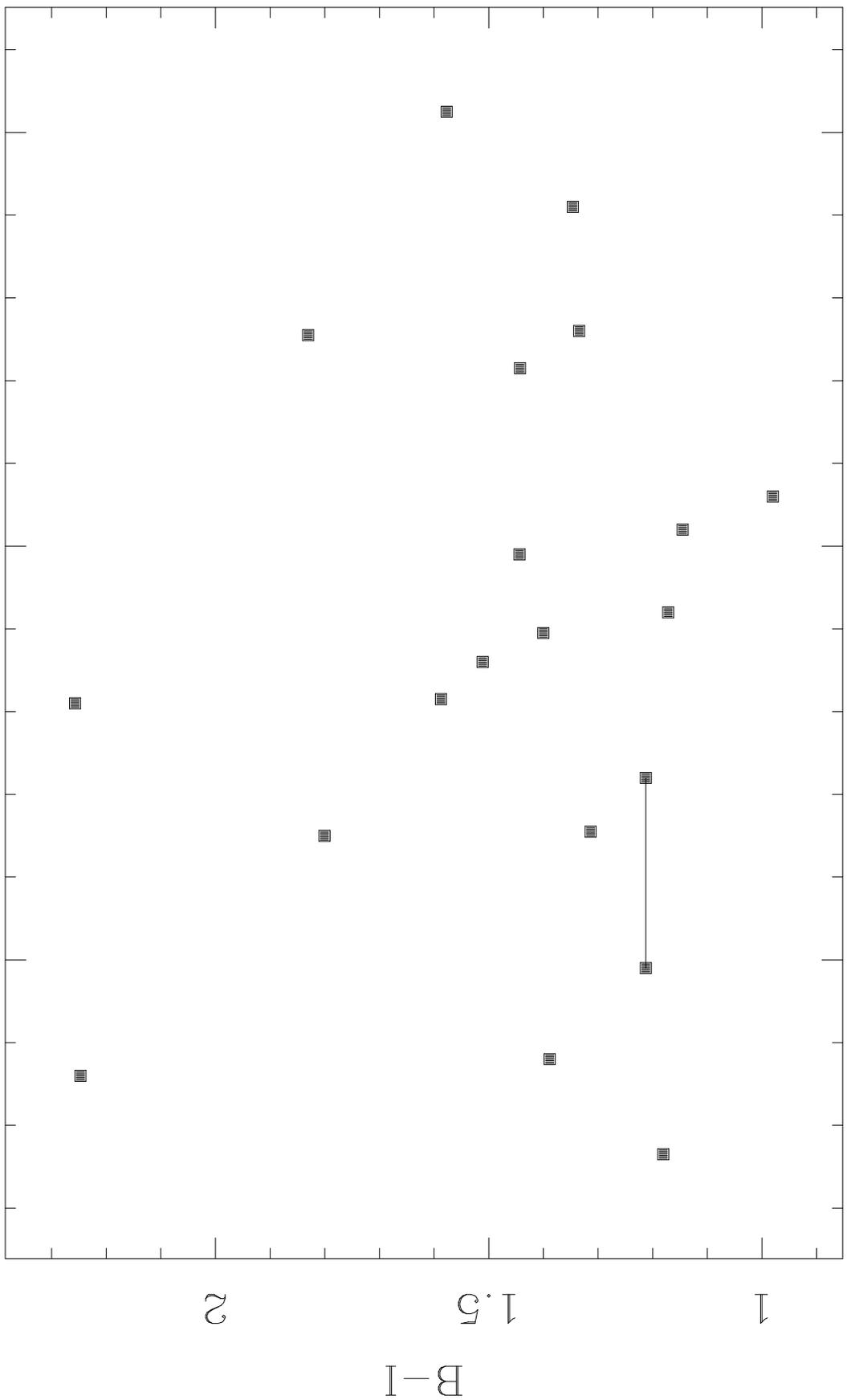

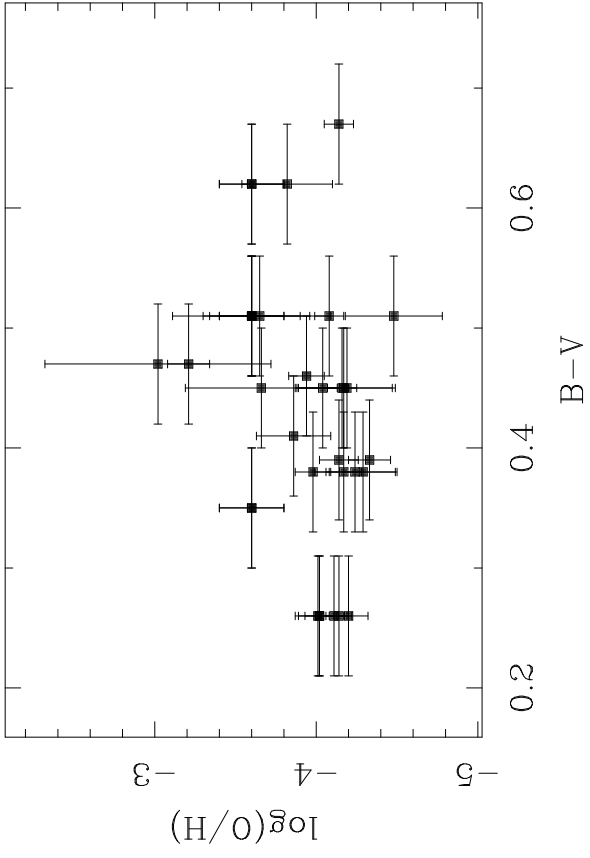
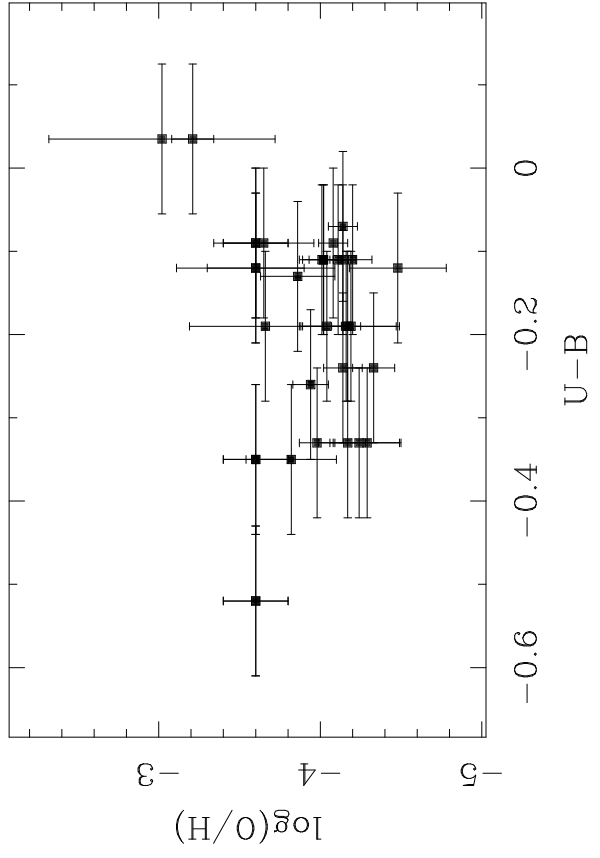
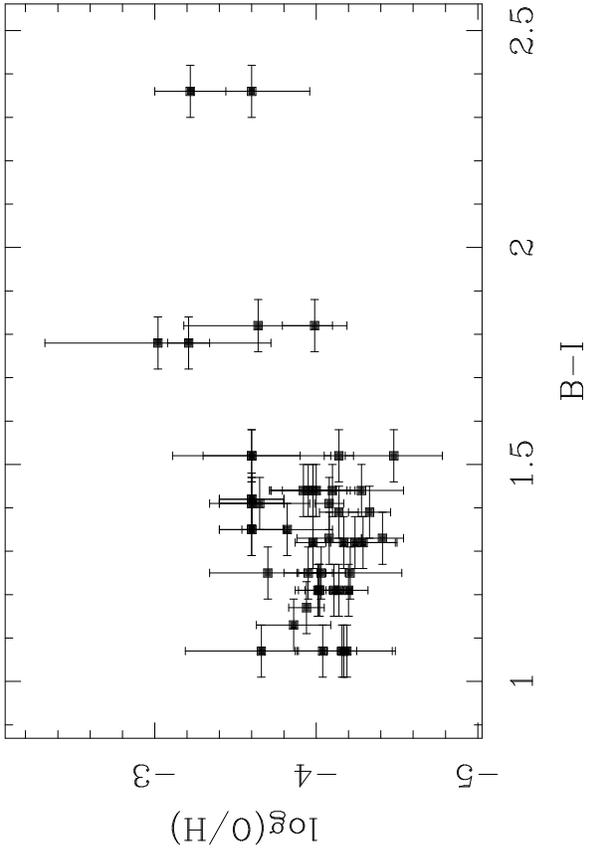
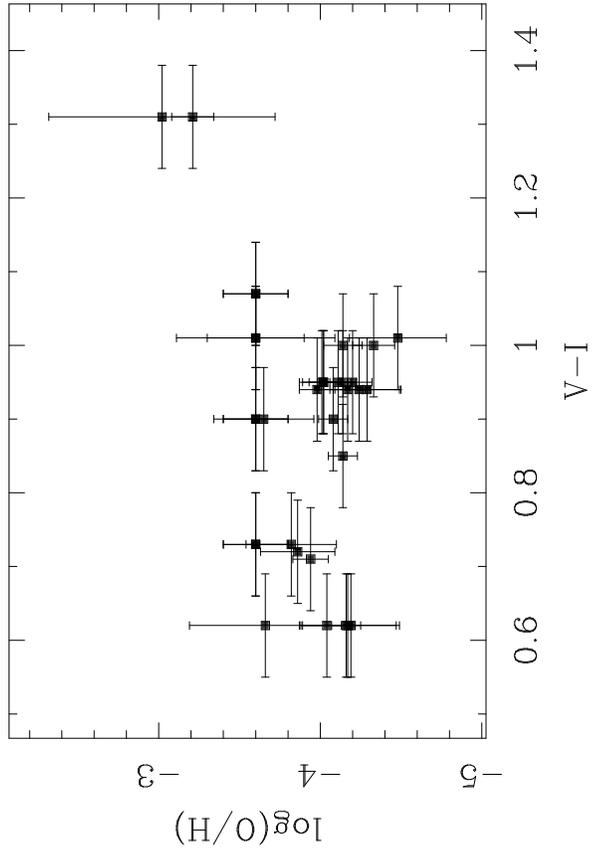

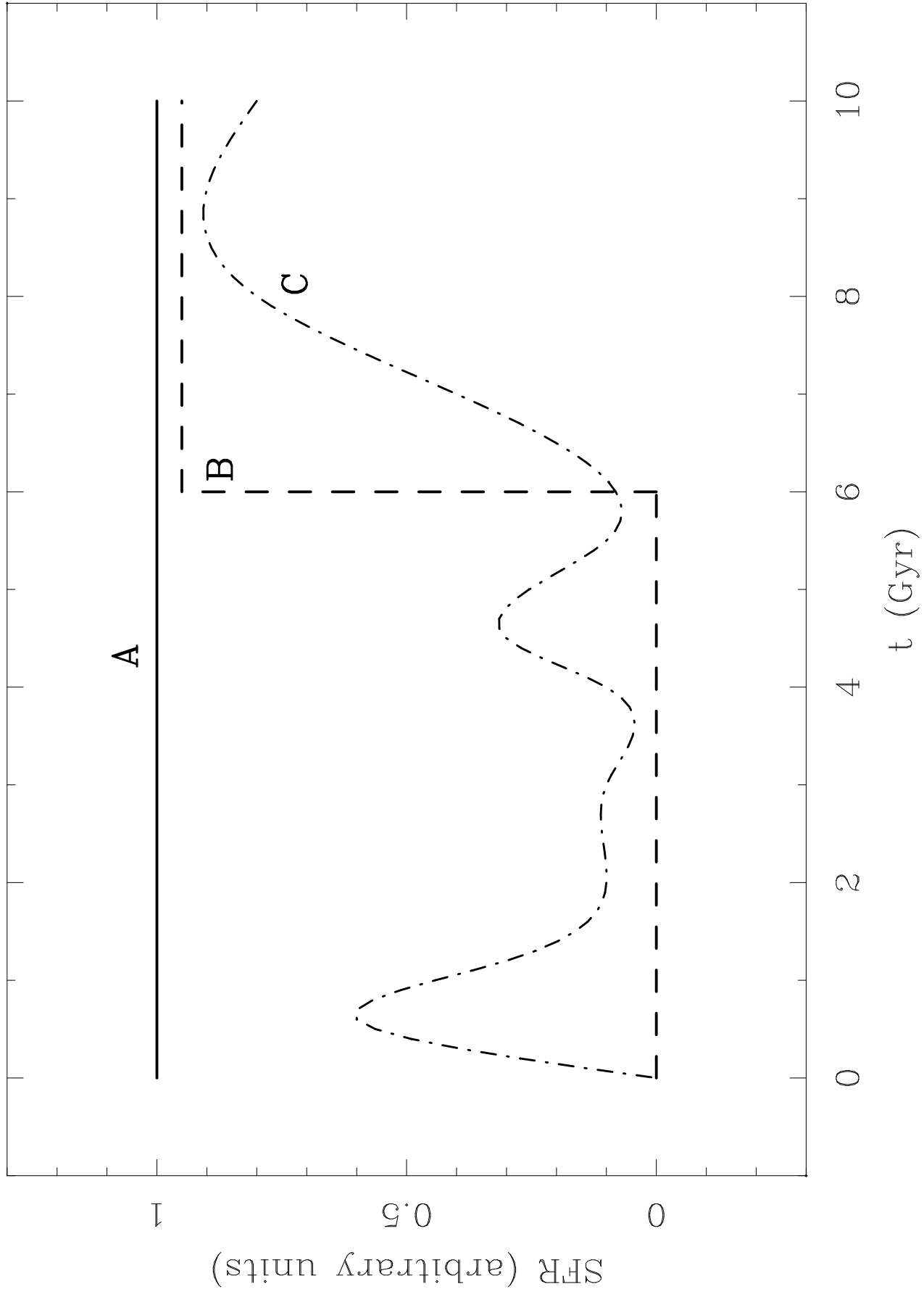

TABLE 1
LSB DISK STRUCTURAL PARAMETERS

| Galaxy | $d_{25}$ | $\alpha$ ($''$) | $\mu_0$ | $\alpha$ (kpc) | $B_{25}$ | $B_T$ | $M_B^T$ | $i$ (°) | $D$ (Mpc) |
|---|---|---|---|---|---|---|---|---|---|
| F415–3 | 50 | 18 | 23.16 | 1.2 | 16.91 | 14.88 | −15.86 | 60 | 14 |
| F469–2 | 43 | 14 | 23.69 | 2.9 | 18.29 | 15.95 | −16.77 | 60 | 35 |
| F530–3 | 39 | 16 | 23.28 | 3.7 | 16.64 | 15.26 | −18.15 | 54 | 48 |
| F561–1 | 41 | 11 | 22.96 | 2.6 | 16.50 | 15.75 | −17.65 | 14 | 48 |
| F563–V1 | 28 | 10 | 23.37 | 1.9 | 17.75 | 16.37 | −16.60 | 48 | 39 |
| F563–V2 | 46 | 7 | 21.95 | 1.6 | 16.48 | 15.72 | −17.59 | 40 | 46 |
| F568–6 | 66 | 23 | 23.38 | 15.8 | 15.28 | 14.57 | −21.17 | 35 | 141 |
| F577–V1 | 25 | ... | ... | ... | 18.49 | ... | ... | 35 | 81 |
| F583–5 | 28 | 9 | 23.48 | 1.4 | 18.53 | 16.70 | −15.85 | 65 | 33 |
| F585–3 | 54 | ... | ... | ... | 16.62 | ... | ... | 74 | 31 |
| F585–V1 | 20 | 9 | 23.40 | 1.1 | 18.40 | 16.62 | −15.33 | 40 | 25 |
| F611–1 | 26 | 13 | 24.24 | 1.5 | 18.39 | 16.67 | −15.14 | 38 | 23 |
| F746–1 | 35 | 8 | 22.18 | 3.0 | 17.54 | 15.66 | −18.80 | 68 | 78 |
| U 1230 | 47 | 16 | 22.88 | 3.0 | 16.20 | 14.85 | −18.06 | 25 | 38 |
| U 5675 | 43 | 19 | 23.70 | 1.5 | 17.47 | 15.30 | −15.73 | 50 | 16 |
| U 5709 | 59 | 13 | 22.49 | 4.1 | 15.97 | 14.92 | −19.15 | 56 | 65 |
| U 6151 | 79 | 24 | 23.21 | 2.1 | 14.90 | 14.30 | −16.92 | 27 | 18 |
| U 6614 | 89 | 37 | 24.28 | 11.9 | 14.48 | 14.43 | −19.67 | 34 | 67 |
| U 9024 | 37 | 35 | 24.47 | 5.6 | 16.67 | 14.74 | −17.84 | 37 | 33 |
| U12695 | 43 | 16 | 23.56 | 4.6 | 17.95 | 15.53 | −18.30 | 49 | 59 |
| " | " | 22 | 24.02 | 6.3 | " | 15.30 | −18.53 | " | " |

TABLE 2
LSB Colors

| Galaxy | $B-V$ | | | $U-B$ | | | $V-I$ | | |
|---|---|---|---|---|---|---|---|---|---|
| | nuc | lum | area | nuc | lum | area | nuc | lum | area |
| F415–3  | 0.37 | 0.46 | 0.52 | −0.47 | −0.26 | −0.12 | 0.68 | 0.71 | ... |
| F469–2  | 0.48 | 0.38 | 0.43 | −0.42 | −0.33 | −0.44 | 0.60 | 0.94 | ... |
| F530–3  | 0.67 | 0.67 | 0.64 | −0.07 | −0.07 | 0.04 | 0.83 | 0.85 | ... |
| F561–1  | 0.64 | 0.41 | 0.44 | 0.03 | −0.13 | −0.12 | 0.88 | 0.72 | 0.71 |
| F563–V2 | 0.54 | 0.51 | 0.36 | 0.04 | −0.12 | 0.02 | 1.10 | 1.01 | 1.21 |
| F577–V1 | 0.44 | 0.35 | 0.37 | −0.20 | −0.52 | −0.31 | 1.08 | 1.07 | 1.08 |
| F611–1  | 0.34 | 0.39 | 0.44 | −0.47 | −0.24 | −0.24 | 0.84 | 1.00 | ... |
| F746–1  | 0.60 | 0.62 | 0.65 | −0.30 | −0.35 | −0.35 | 0.70 | 0.73 | ... |
| U 1230  | 0.62 | 0.45 | 0.42 | −0.05 | −0.19 | −0.18 | 0.81 | 0.62 | 0.56 |
| U 5709  | 0.83 | 0.47 | 0.48 | 0.13 | 0.03 | 0.04 | 1.34 | 1.31 | 1.35 |
| U 6151  | 0.51 | 0.51 | 0.51 | −0.02 | −0.09 | −0.07 | 1.08 | 0.90 | 0.89 |
| U 6614  | 1.01 | 0.72 | 0.53 | 0.39 | 0.25 | 0.32 | 1.53 | 1.60 | 1.72 |
| U12695  | 0.43 | 0.26 | 0.37 | −0.19 | −0.11 | −0.17 | 0.82 | 0.95 | ... |

TABLE 3
$B - I$ COLORS

| Galaxy | $B - I$ | | |
| --- | --- | --- | --- |
| | nuc | lum | area |
| F568-6 | 2.77 | 2.36 | 2.26 |
| F583-5 | 1.28 | 1.33 | ... |
| F585-3 | 1.32 | 1.44 | 1.44 |
| U 5675 | 1.77 | 1.82 | 1.80 |
| U 9024 | 1.51 | 1.25 | 1.18 |

# FIGURE CAPTIONS

**Figure 1a.** The $B$ radial surface brightness profiles of LSB galaxies (squares), together with exponential fits to the disk components (lines). The abscissa is the radius along the major axis.

**Figure 1b.** The profile of F568–6 is in good agreement with that reported by Bothun et al. (1990). F577–V1 and F585–3 are not well fit by exponential profiles.

**Figure 1c.** The surface brightness profile of UGC 6614 is in good agreement with that obtained with the OREAD (Aldering & Bothun 1991) focal reducing camera (open circles). Only points with $r > 50''$ are included in the fit to the exponential disk.

**Figure 1d.** UGC 12695 is acceptably fit by two profiles, the one with the lower central surface brightness perhaps being consistent with the presence of a second, short scale length exponential disk as the central concentration.

**Figure 2.** The deficit in brightness measured by the light within the $\mu = 25$ mag arcsec$^{-2}$ isophote relative to that from the integration of the exponential profile. Note that $B_{25}$ substantially underestimates the total flux (by a factor of $\sim 4$ in most cases). The isophotal magnitude falls well below the line of equality despite including light from any bulge component which is *not* included in the exponential fit. Indeed, the only points which approach the line of equality are due to galaxies with prominent bulges. The large disparity shows how easy it is to underestimate the total amount of light emitted by LSB galaxies, and the importance of using profile fits in evaluating size and luminosity — isophotal diameters such as $d_{25}$ are similarly inadequate measures of the sizes of LSB disks.

**Figure 3.** Histograms of the structural parameters (a) $\mu_0$ and (b) $\alpha$ (in kpc). Note that there is no preferred value of either.

**Figure 4.** The observed structural parameters $(\mu_0, \alpha)$ plotted against one another. The solid squares are the galaxies in this study. The line connects the two acceptable fits for UGC 12695. For comparison, the LSB sample of Romanishin et al. (1983;

open squares) and the HSB sample of van der Kruit (1987; open triangles) are also plotted. The size of typical error bars is shown by the cross in the corner. Disk galaxies cover a wide range in both structural parameters with no apparent correlation, though there may be a lack of very large HSB disks.

**Figure 5.** The central surface brightness plotted against the absolute magnitude of the disk component. Symbols as per Figure 4. An apparent correlation is induced by the calculation of the total disk luminosity from equation 2, and may be an artifact arising from incomplete sampling of the $(\mu_0, M_B)$ plane. No correlation is apparent in the present dataset, and LSB galaxies can be very luminous. Very roughly, this diagram traces the initial conditions of galaxy formation, with luminosity corresponding to mass and surface brightness to collapse epoch (see text).

**Figure 6.** LSB galaxy colors. (a) $U - B$ vs. $B - V$ and (b) $B - V$ vs. $V - I$. The area weighted colors are used when available; luminosity weighted colors are used when not. Though mostly blue, LSB disks do cover a substantial range of color. As with the structural parameters, no correlations are evident, though there may be a tendency for disks which are blue in $B - V$ to be red in $V - I$. The centroid of the $(U - B, B - V)$ colors is near to that of the constant star formation rate model of Larson & Tinsley (1978).

**Figure 7.** $B - I$ vs. (a) $\mu_0$ and (b) $\alpha$. No correlation exists between this (or any other) color and any structural parameter. The lack of a trend in (a) completely rules out evolutionary fading as the cause of the low surface brightnesses of these disks.

**Figure 8.** The H II region oxygen abundances of these LSB galaxies (McGaugh 1993) plotted against the various colors. That no clear trends are visible means that the metallicity is *not* the primary cause of the blue colors, though it may contribute at a level which is drowned out by differences in the star formation histories.

**Figure 9.** Simple star formation histories. The star formation rate is constant in both scenarios A and B, with star formation commencing at some later epoch in B. The current data do not discriminate between B and some more complicated form

like C, but do seem to indicate that on average the stellar populations in LSB galaxies are weighted towards youth.